# The Impact of Human Factors on the Participation Decision of Reviewers in Modern Code Review

**Shade Ruangwan · Patanamon Thongtanunam**
**Akinori Ihara · Kenichi Matsumoto**

**Abstract** Modern Code Review (MCR) plays a key role in software quality practices. In MCR process, a new patch (i.e., a set of code changes) is encouraged to be examined by reviewers in order to identify weaknesses in source code prior to an integration into main software repositories. To mitigate the risk of having future defects, prior work suggests that MCR should be performed with sufficient review participation. Indeed, recent work shows that a low number of participated reviewers is associated with poor software quality. However, there is a likely case that a new patch still suffers from poor review participation even though reviewers were invited. Hence, in this paper, we set out to investigate the factors that are associated with the participation decision of an invited reviewer. Through a case study of 230,090 patches spread across the Android, LibreOffice, OpenStack and Qt systems, we find that (1) 16%-66% of patches have at least one invited reviewer who did not respond to the review invitation; (2) human factors play an important role in predicting whether or not an invited reviewer will participate in a review; (3) a review participation rate of an invited reviewers and code authoring experience of an invited reviewer are highly associated with the participation decision of an invited reviewer. These results can help practitioners better understand about how human factors associate with the participation decision of reviewers and serve as guidelines for inviting reviewers, leading to a better inviting decision and a better reviewer participation.

Shade Ruangwan, Kenichi Matsumoto
Nara Institute of Science and Technology, Japan
E-mail: {shade.ruangwan.sj1,matumoto}@is.naist.jp

Akinori Ihara
Wakayama University, Japan
E-mail: ihara@sys.wakayama-u.ac.jp

Patanamon Thongtanunam
The University of Adelaide, Australia
E-mail: patanamon.thongtanunam@adelaide.edu.au

## 1 Introduction

Code review is one of the well-known software quality practices in software development process (Huizinga and Kolawa, 2007, p. 260). The main motivation of code review is to early identify defects in source code before a software product is released (Ackerman et al, 1989; Bacchelli and Bird, 2013). Traditionally, code review is a formal and well-documented code inspection process which is performed by well-allocated team members (Fagan, 1976).

Nowadays, many modern software development teams tend to adopt a lightweight variant of code review called Modern Code Review (MCR) (Beller et al, 2014). Broadly speaking, for every new patch (i.e., a set of code changes), a patch author invites a set of reviewers (i.e., team members) to examine the patch prior to an integration into main software repositories. MCR tends to focus on collaboration among team members to achieve high quality of software products. Such practices of MCR also provide additional benefits to team members such as knowledge transfer and increasing team awareness (Bacchelli and Bird, 2013).

However, the lightweight variant of MCR are prone to lower review participation than the formal code review due to its informal nature. Instead of carefully assigning reviewers like the formal code review (Fagan, 1976, 1986), reviewers of MCR can decide whether or not to participate a review. Hence, the review participation becomes one of the main challenges in MCR process. Several studies find that a number of participated reviewers has an impact on software quality and code review timeliness (Bavota and Russo, 2015; Bettenburg et al, 2015; McIntosh et al, 2014). Moreover, Kononenko et al (2015) find that the number of invited reviewers have a statistically significant impact on review bugginess.

Finding reviewers in geographically-distributed software development teams can be difficult (Thongtanunam et al, 2015b). Several studies propose an approach to help patch authors find reviewers who will participate in a review. The common intuition of the prior work is that a reviewer familiars with the code on the patch is more likely to give a better review than others (Balachandran, 2013; Thongtanunam et al, 2015b; Xia et al, 2015; Yu et al, 2014; Zanjani et al, 2016).

In addition to finding reviewers, recent work finds that a patch author should prepare a small size of a patch, provide a descriptive subject, and explain change log message to increase the likelihood of receiving review participation (Thongtanunam et al, 2016a). While prior studies have explored several technical factors (i.e., reviewer experience and patch characteristics) that share a link to review participation, no prior study confirms a link between the human factors and the participation decision of a reviewer.

In this paper, we analyze descriptive statistics of reviewers who did not respond to the review invitation to understand the current practices of the participation decision of reviewers. To better understand the signals that can relate to the participation decision of a reviewer, we construct statistical models that predict the participation decision of reviewers (i.e., whether or not an invited reviewer will participate in a review) using the nonlinear logistic regression modeling technique. The nonlinear logistic regression modeling technique allows us not only to predict an outcome of interest, but also to explore the relationships between independent variables and dependent variable (Harrell Jr., 2002). In particular, we construct two prediction models for each studied dataset. One is our proposed model that uses human factors, reviewer experience, and patch characteristics. The other one is the baseline model that uses only reviewer experience and patch characteristics. Through a case study of 230,090 patches spread across the Android, LibreOffice, OpenStack and Qt systems, we address the followings research questions:

**(RQ1) How often do patches suffer from the unresponded review invitations?**
We find that 16%-66% of patches have at least one invited reviewer who did not respond to the review invitation. Moreover, the number of invited reviewers shares a positive correlation with the number of reviewers who did not respond to the review invitation.

**(RQ2) Can human factors help determining the likelihood of the participation decision of reviewers?**
Our proposed prediction models, which include human factors, achieve an AUC value of 0.82-0.89, a Brier score of 0.06-0.13, a precision of 0.68-0.78, a recall of 0.24-0.73, and an F-measure of 0.35-0.75. Moreover, we find that including human factors to the baseline models, that use only reviewer experience and patch characteristics, increases their F-measure by 17%-1,800%. These results suggest that human factors play an important role in determining the likelihood of the participation decision of reviewers.

**(RQ3) What are the factors mostly associated with participation decision?**
We find that in addition to reviewer experience, human factors (e.g., the review participation rate of an invited reviewer) are also highly associated with the likelihood that an invited reviewer will participate in a review.

Our results lead us to conclude that patches undergoing MCR process often suffer from the unresponded review invitations. Human factors of invited reviewers play a crucial role in the participation decision. Both technical and human factors should be considered when determining the likelihood that an invited reviewer will participate in a review. In addition to the experience of invited reviewers, these findings highlight the need of considering the human factors before inviting reviewers in order to increase review participation.

### 1.1 Paper organization

The remainder of this paper is organized as follows. Section 2 situates our three research questions with respect to the related works. Section 3 describes MCR process. Section 4 describes our case study design, while Section 5 presents the results with respect to our three research questions. Section 6 presents our survey study. Section 7 discusses our findings. Section 8 discloses the threats to the validity of our study. Finally, Section 9 draws conclusions.

## 2 Background and Research Questions

To create and deliver high quality software products, software development processes require a strong collaboration among software developers (Whitehead, 2007).

However, a collaboration is challenging, especially for the geographically-distributed teams like Open Source Software (OSS) projects (Lanubile et al, 2010). Hahn et al (2008) also report that one of the common reasons for the failure in OSS projects is the poor collaborations in software development teams.

Code review is one of the software development processes that require an intensive collaboration among software developers. Code review is a code examination process to improve software quality (Huizinga and Kolawa, 2007, p. 260). Specifically, the main motivations of code review are to find and remove software defects early in the development cycle (Bacchelli and Bird, 2013). In addition, several studies also find that code review can improve software security (Edmundson et al, 2013; McGraw, 2004), reduce code complexity, increase code readability, and reduce a risk of inducing bug fixes (Bavota and Russo, 2015).

Code review can be performed through either formal or lightweight processes (Huizinga and Kolawa, 2007, p. 260). The formal code review process involves well-defined steps which are carried out by face-to-face meeting (Fagan, 1976). Moreover, before beginning the code review process, documents and participants for the meeting are carefully prepared. On the other hand, the lightweight code review process, also known as Modern Code Review (MCR), is less formal. The process does not require a face-to-face meeting. Instead, MCR is perform through online tools such as Gerrit,[1] Review Board,[2] and Crucible.[3] Such a process facilitates collaboration in teams, especially for the geographically-distributed software projects (Meyer, 2008). Nowadays, MCR has been widely used in many software development organizations such as Android Open Source Project,[4] Eclipse Foundation,[5] and Mozilla.[6]

Despite the ease of performing code review of MCR, participation becomes a challenge since developers can decide whether or not to participate a review. For example, Rigby and Storey (2011) report that a reviewer may not participate in a review of a patch since it is not in the reviewer interest. Prior work also reports that participation in MCR is associated with the quality of the code review process (Kononenko et al, 2015).

Several studies have investigated the impact that poor review participation can have on software quality and code review process. Bavota and Russo (2015) find that a patch with lower review participation has a higher chance of inducing bug fixes. McIntosh et al (2014) find that the lack of review participation has a negative impact on software quality. More specifically, they find that the more often that the components are reviewed with low reviewer participation, the more post-release defects the components contain. Thongtanunam et al (2015a) find that a file that involves fewer reviewers when undergoes code review is more likely to be found defective later on. In addition to the software quality, Bettenburg et al (2015) find that the overall review time and the delay of receiving first feedback increase when the ratio of reviewers to patch authors in the project decreases. While several studies have uncovered the impact of review participation,

[1] `https://www.gerritcodereview.com/`
[2] `https://www.reviewboard.org/`
[3] `https://www.atlassian.com/software/crucible/`
[4] `https://android-review.googlesource.com/`
[5] `https://git.eclipse.org/`
[6] `https://reviewboard.mozilla.org/`

little is known about how often poor review participation occurs in MCR process. Moreover, understanding the current practices of reviewer participation would help teams to increase an awareness of the poor review participation as well as to better manage the code review process. Hence, we perform an exploratory study to investigate how often do reviewers decide to not respond to the review invitation and we set out to address the following research question:

**RQ1: How often do patches suffer from the unresponded review invitations?**

To help patch authors find the most knowledgeable reviewer to better review the patch, several studies propose an approach to recommend reviewers who will participate in a review. Balachandran (2013) proposes an algorithm which uses a change history of source code in lines that reviewers have reviewed in the past. Yu et al (2014) compute a reviewer expertise and common interests between patch authors and reviewers using textual semantic of pull requests and comment network between patch authors and reviewers in GitHub. Thongtanunam et al (2015b) propose RevFinder which computes the file path similarity between a new patch and prior patches that reviewers have reviewed. Xia et al (2015) propose Tie which is based on the textual content of the patch, file path, and patch submit time. Zanjani et al (2016) propose cHRev which uses the historical contributions of reviewers. One common intuition of these studies is that a patch author should invite reviewers based on the past experience. In other words, a reviewer is more likely to review if the reviewer is familiar with that area of code in the patch.

While the reviewer related experience is known to link with review participation (Balachandran, 2013; Thongtanunam et al, 2015b; Xia et al, 2015; Yu et al, 2014; Zanjani et al, 2016), others factors may also associate with the review participation. Rigby and Storey (2011) find through interviews that time, priorities and interest of the invited reviewers are the main reasons why they do not participate in a broadcast based peer review. Recently, Thongtanunam et al (2016a) have investigated whether technical factors like patch characteristics can lead to poor review participation. However, little is known whether non-technical factors like human factors can be associated with the participation decision of reviewers. To explore the impact human factors on the review participation and help patch authors in inviting reviewers, we set out to address the following research question:

**RQ2: Can human factors help determining the likelihood of the participation decision of reviewers?**

Having investigated the extent of how often patches suffer from the poor review participation and the importance of human factors, a better understanding of the factors associated with participation decision would help software development teams to develop better strategies for the code review process. For example, software development teams can create a guideline, especially for new developers, on best practices in inviting reviewers and getting reviewed, because not receiving a response and learning the process are reported to be the barriers for new developers (Lee et al, 2017; Steinmacher et al, 2015). Several studies have investigated the impact of factors that are recorded during the code review process. Baysal et al (2013) find that organizational and personal factors can have an impact on the review timeliness and the likelihood that a patch will be accepted. Rigby et al (2014) find that the number of reviewers and the size of the patches can have an impact on the review timeliness and effectiveness. Bosu and Carver (2014) find that patch author reputation can have an impact on the first feedback interval, review interval, and patch acceptance rate. Armstrong et al (2017) find that code review medium (i.e., broadcast or unicast based peer review) can have an impact on the review effectiveness and quality. Hence, to better understand the impact that human factors, reviewer experience, and patch characteristics can have on the participation decision of reviewers, we address the following question:

**RQ3: What are the factors mostly associated with participation decision?**

## 3 Modern Code Review

Modern Code Review (MCR) is a lightweight variant of code inspection process that is often supported by tools. MCR has been widely adopted by many software development organizations (Rigby et al, 2012). Main purposes of MCR include detecting and fixing defects earlier in the software development cycle (Bacchelli and Bird, 2013). Figure 1 provides an overview of MCR process, which is based on Gerrit code review. Gerrit[7] is a web-based code review tool that tightly integrates with Git version control system. Below, we describe the MCR process of our studied systems.

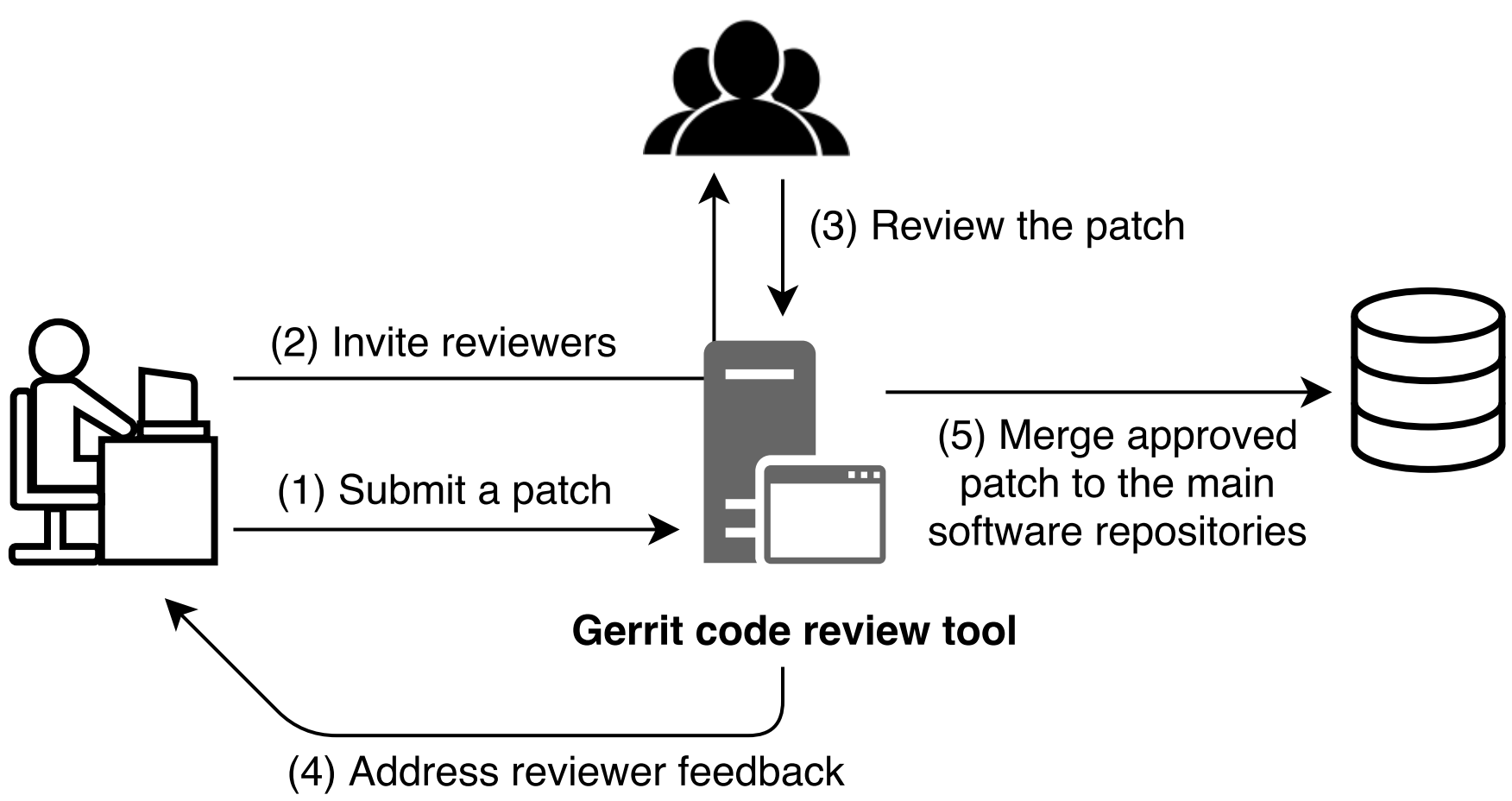

Fig. 1: The overview of MCR process.

**(1) Submit a patch.** When a patch author (i.e., a developer) makes a new patch, i.e., a set of code changes, the new patch is examined prior to an integration

[7]https://www.gerritcodereview.com/

to the main software repositories. Hence, the patch author uploads the new patch to Gerrit.

**(2) Invite reviewers.** A patch author invites reviewers through Gerrit. Then, Gerrit will notify the invited reviewers by email and add the review task to the reviewing list. Reviewers can decide to respond or not respond to the review invitation.

**(3) Review the patch.** The invited reviewers can respond to the review invitation by inspecting the patch and providing feedback in a comment section. To indicate whether the patch should be integrated into main software repositories, the reviewers can also provide a review score ranging from +2 to -2. A review score of +1 indicates that the reviewers agree with the patch, however, they need a confirmation from other reviewers. A review score of +2 indicates that the patch can be integrated into the main software repositories. Similarly, a review score of -1 indicates that the reviewers prefer a revision of the patch before an integration into the main software repositories. A review score of -2 indicates that the patch requires a major revision.

**(4) Address reviewer feedback.** Once the reviewers have reported potential problems, the patch author revises the patch according to the reviewers' feedback. Then, the revised patch is submitted to the Gerrit system in order to be re-inspected by the reviewers. Thus, creating a feedback cycle.

**(5) Merge approved patch to the main software repositories.** Once the review of the patch has reached a decision and the patch receives a review score of +2, the patch is integrated into main software repositories and marked as `merged` in Gerrit. The patch is marked as `abandoned` if the reviewers evaluate that patch does not meet a sufficient quality level or require too much rework.

## 4 Case Study Design

In this section, we describe the studied systems, data preparation, and analysis approaches. Figure 2 provides an overview of our case study design.

### 4.1 Studied Systems

To address our research questions, we select large software systems that actively use modern code review. Since we will analyze the participation decision of reviewers, we need to ensure that the review participation is mainly recorded in the code review tools. Therefore, we select to study the code review process of Android, LibreOffice, OpenStack, and Qt systems since these systems have a large number of patches that have been recorded in the code review tool (see Table 1). Android[8] is an open source mobile operating system developed by Google. LibreOffice[9] is an open source office suite developed by The Document Foundation. OpenStack[10] is an open source cloud operating system started by Rackspace Hosting and NASA and currently managed by the OpenStack Foundation. Qt[11] is an open source cross-platform application framework developed by The Qt Company.

[8] `https://source.android.com/`
[9] `https://www.libreoffice.org/`
[10] `https://www.openstack.org/`

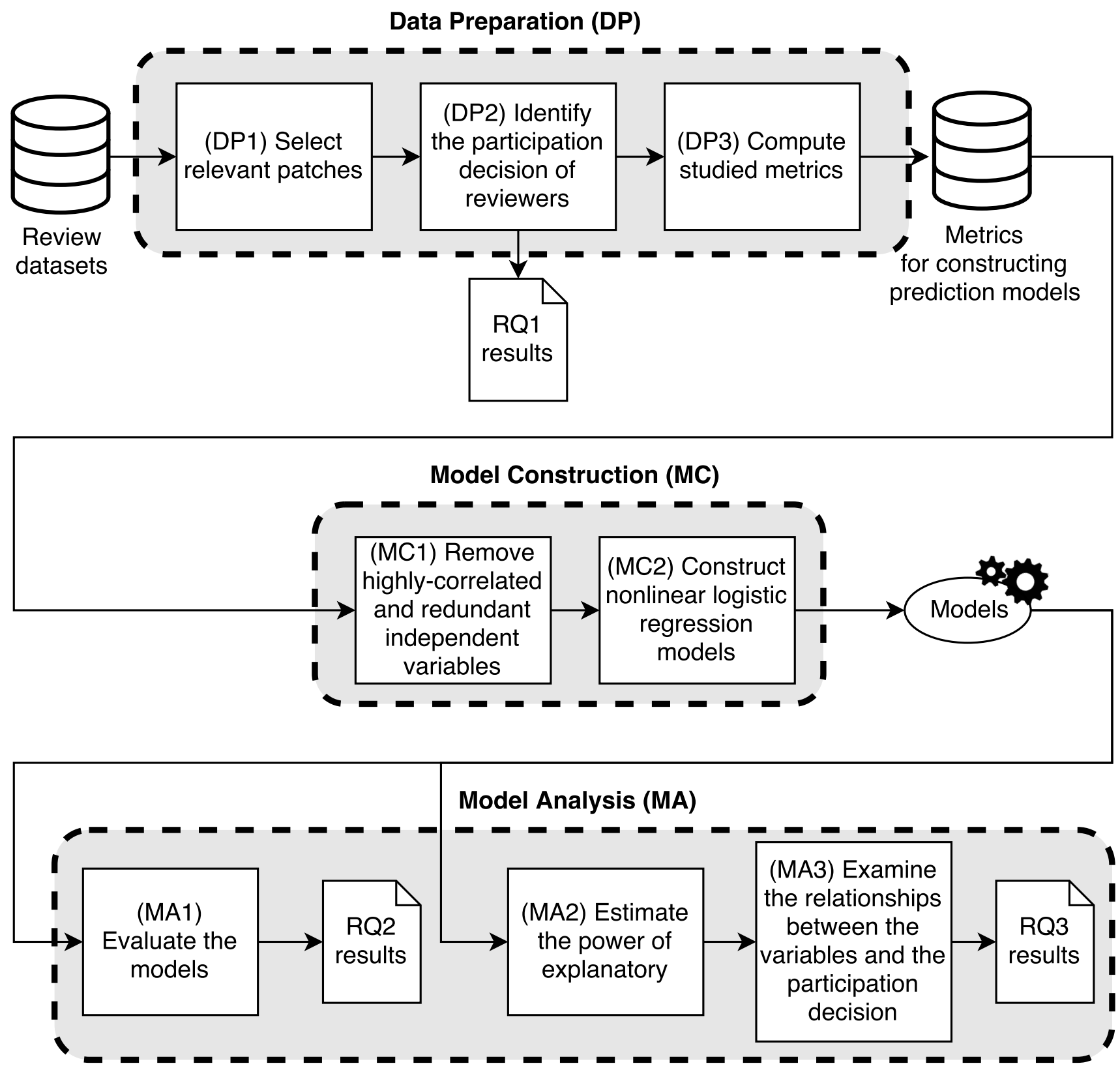

Fig. 2: An overview of our case study design.

For Android, OpenStack, and Qt systems, we use review datasets of Hamasaki et al (2013) which are often used in prior studies (McIntosh et al, 2014; Thongtanunam et al, 2015b). For LibreOffice system, we use a review dataset of Yang et al (2016a). The datasets include patch information, review discussion, and developer information. To retrieve a complete list of invited reviewers and review scoring information, we use REST API that is provided by Gerrit.[12] Table 1 provides a statistical summary of the review datasets.

## 4.2 Data Preparation (DP)

Before performing an empirical study, we prepare the studied datasets. Figure 2 shows an overview of our data preparation approach which consists of three main steps: (DP1) select relevant patches, (DP2) identify the participation decision of reviewers, and (DP3) compute studied metrics. We describe each step below.

[11] https://www.qt.io/
[12] https://gerrit-review.googlesource.com/Documentation/rest-api.html/

Table 1: Summary of the studied datasets.

| | Android | LibreOffice | OpenStack | Qt |
|---|---|---|---|---|
| Start Time | 10/2008 | 3/2012 | 7/2011 | 5/2011 |
| End Time | 12/2014 | 11/2016 | 12/2014 | 12/2014 |
| Duration | 6 Years | 4 Years | 3 Years | 4 Years |
| #Patches | 36,771 | 18,716 | 108,788 | 65,815 |
| Avg. #Patches/Years | 6,129 | 4,679 | 36,263 | 16,454 |
| #Reviewers | 2,049 | 410 | 3,734 | 1,238 |
| #Patch Authors | 1,428 | 557 | 3,249 | 1,011 |

**(DP1) Select relevant patches.** In this study, we only study patches that have been masked as either `merged` or `abandoned`. We exclude patches with the `open` status from the studied datasets because the participation decision of reviewers of the `merged` and `abandoned` patches have been made and there is a less likely case that the invited reviewers continue to participate the reviews.

Furthermore, we remove patches that have only the patch author who is in the list of invited reviewers apart from automated checking systems (i.e., self-reviewed patches) since such kind of patches intuitively do not have participated reviewers. We classify a patch where its description contains "merge branch" or "merge" as VCS bookkeeping activities (e.g., branch-merging patches) and remove them since these patches are involved with prior patches that were already reviewed. Moreover, we observe that on average, 44%-78% of these VCS bookkeeping patches are self-reviewed patches.

**(DP2) Identify the participation decision of reviewers.** Before identifying the participation decision of reviewers, we remove the accounts of automated checking systems (e.g., Jenkins CI or sanity checks) from the list of invited reviewers, since these systems will automatically run a check on every patch. We identify the accounts of automated checking systems of Android system as suggested by (Mukadam et al, 2013). We use an account list of automated checking systems of OpenStack and Qt systems, which is provided by a prior study (Thongtanunam et al, 2016a). We identify an account list of automated checking systems of LibreOffice based on comments that are posted to patches. In particular, we manually identify an account that repeatedly posts messages that contain "Build Started" or "Build Failed" keywords since these keywords indicate the process status of the automated checking systems.

Once we remove the accounts of automated checking systems, we identify the participation decision of reviewers. We identify an invited reviewer who did not participate in the review by providing neither a review score nor comments as a reviewer who did not respond to the review invitation. We identify the remaining invited reviewers as a reviewer who responded to the review invitation. Figure 3 shows an example of identifying participation decision of invited reviewers, where a patch author (i.e., A) invites 3 reviewers (i.e., reviewers R1, R2, and R3). In this example, reviewer R2 is identified as a reviewer who did not respond to the review invitation as reviewer R2 did not provide a review score nor feedback. Reviewers

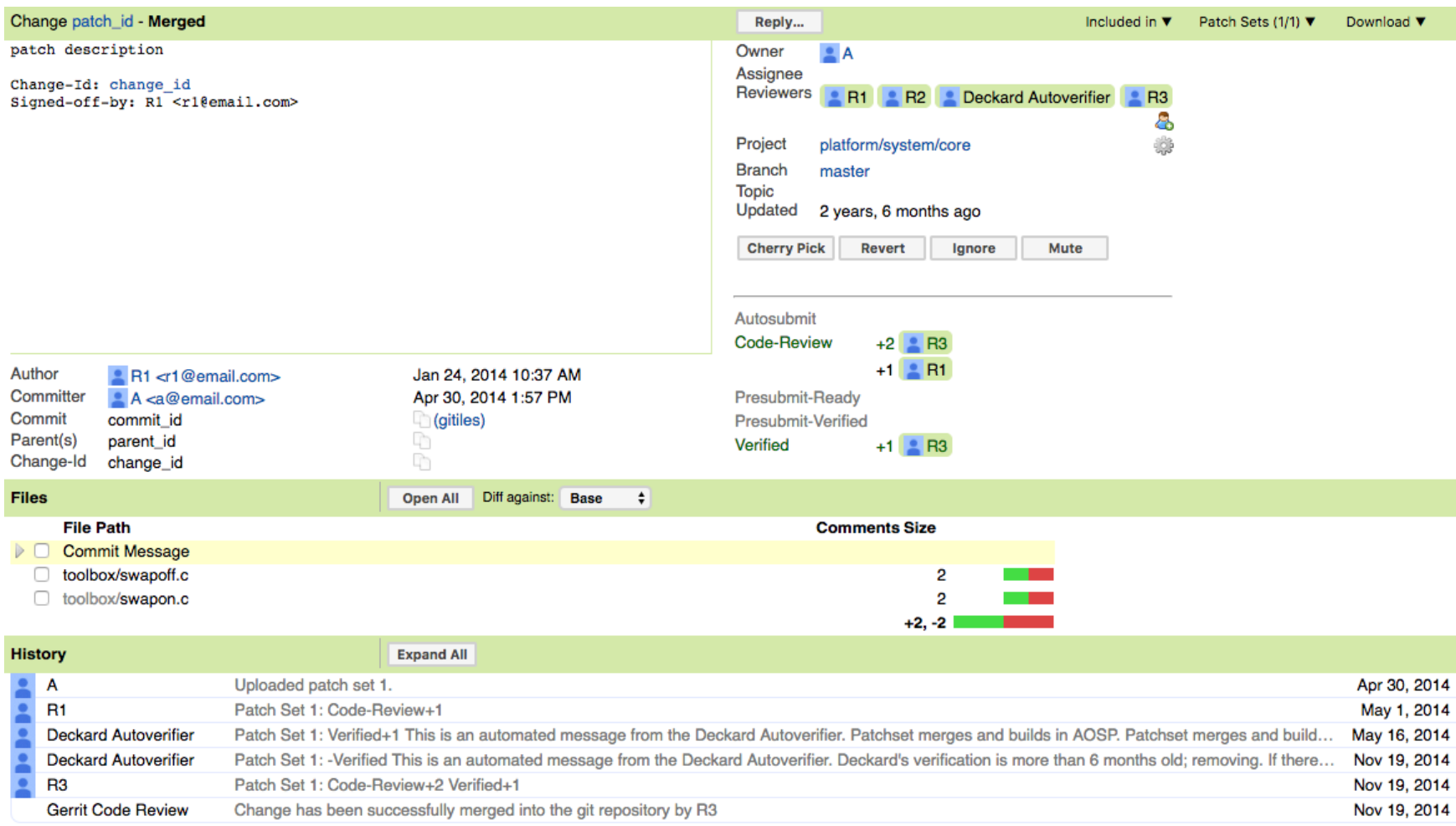

Fig. 3: An example of identifying participation decision of reviewers.

R1 and R3 are identified as a reviewer who responded to the invitation since they provide a review score.

**(DP3) Compute studied metrics.** To understand the impact of human factors on the participation decision of reviewers, we extract 12 metrics from the datasets. The metrics are grouped along 3 dimensions; i.e., human factors, reviewer experience, and patch characteristics. Table 2 describes the conjecture and rationale of each metric. Below, we describe the calculation of our metrics.

***Human Factors Dimension*** Human factors dimension measures reviewer related environment and reviewer past activities. Human factors dimension is divided into two sub-dimensions:

*Review Workload* Review workload measures review workload of an invited reviewer at the time when the invited reviewer received a new review invitation. *Number of Concurrent Reviews* counts how many patches that an invited reviewer was reviewing at the time when the studied patch is created. We consider an invited reviewer was reviewing a patch when the invited reviewer had provided a review score or comments to that patch. We also count only patches that have not reached a final decision (i.e., have not been marked as `merged` or `abandoned`) at the time when the studied patch is created. Figure 4 shows an example of counting the number of concurrent reviews where reviewers A and B were an invited reviewer of the studied Patch #1. In this example, the number of concurrent reviews of reviewer A is 2 (i.e., Patches #2, and #3). However, reviewer B does not have any concurrent reviews since he did not participate in Patches #2 and #3. Furthermore, in the example, Patch #4 will not be considered for reviewers A and B since Patch #4 was created after the studied patch. *Number of Remaining Reviews* counts the number of patches where an invited reviewer was invited, yet the invited reviewer did not participate in at the time when the studied patch is created. Similar to the number of concurrent reviews, we count only patches that were created before the studied patch but had not reached a final decision at the time when the studied patch is created. Using the same example in Figure 4, reviewer A has no remaining review while reviewer B has 2 remaining reviews (i.e., Patches #2, and #3).

Table 2: The studied metrics

| Metric | Conjecture | Rationale |
|---|---|---|
| ***1 Human Factors Dimension*** | | |
| *1.1 Review Workload* | | |
| Number of Concurrent Reviews | The more concurrent review tasks the invited reviewer has, the more likely that the invited reviewer will not respond to a new review invitation | A reviewer who is burdened with large number of review tasks may not have time to review a new patch. |
| Number of Remaining Reviews | The more remaining review tasks the invited reviewer has, the more likely the invited reviewer will not respond to a new review invitation. | |
| *1.2 Social Interaction* | | |
| Familiarity between the Invited Reviewer and the Patch Author | The invited reviewer is more likely to respond to a new review invitation if the reviewer reviewed the prior patches of the patch author before. | A reviewer may prefer to review patches of the patch author who the invited reviewer knows. |
| Median Number of Comments | The more comments that the invited reviewer had provided in the past, the more likely that the invited reviewer will respond to a new review invitation. | A large number of comments that the reviewer has provided in the past may indicate that the reviewer is active in the system. |
| Review Participation Rate | A reviewer with a high rate of review participation is more likely to respond to review invitation. | A high rate of review participation may indicate that the reviewer is active in the system. |
| Number of Received Review Invitations | A reviewer who received many review invitations is more likely to respond to a review invitation. | Such a reviewer may be an expert who is widely known by the patch authors. |
| Core Member Status | A core reviewer is more likely to respond to a new review invitation | Core reviewers may be more active than non-core reviewers (Vasilescu et al, 2014). It is also possible that most of the activities may be carried out by a group of core reviewers (Goeminne and Mens, 2011). |
| ***2 Reviewer Experience Dimension*** | | |
| Reviewer Code Authoring Experience | The invited reviewer is more likely to respond to a new review invitation of a patch that the reviewer has related authoring experience. | A reviewer may prefer to review new patches that the reviewer has related experience (Liang and Mizuno, 2011; Thongtanunam et al, 2015b; Xia et al, 2015). |
| Reviewer Reviewing Experience | The invited reviewer is more likely to respond to a new review invitation of a patch that the reviewer has related reviewing experience. | |
| ***3 Patch Characteristics Dimension*** | | |
| Patch Size | A patch with small code changes is more likely to get a review. | Some initial qualitative evidences indicate that small patches are easier to review than large patches (Mishra and Sureka, 2014; Rigby et al, 2014). |
| Patch Author Code Authoring Experience | The invited reviewer is likely to respond to a new review invitation if the patch author is an experienced developer. | An experienced patch author may widely known for his/her capability which encourage reviewers to work with (Bosu and Carver, 2014). |
| Patch Author Reviewing Experience | The invited reviewer is likely to respond to a new review invitation if the patch author is an experienced reviewer. | |

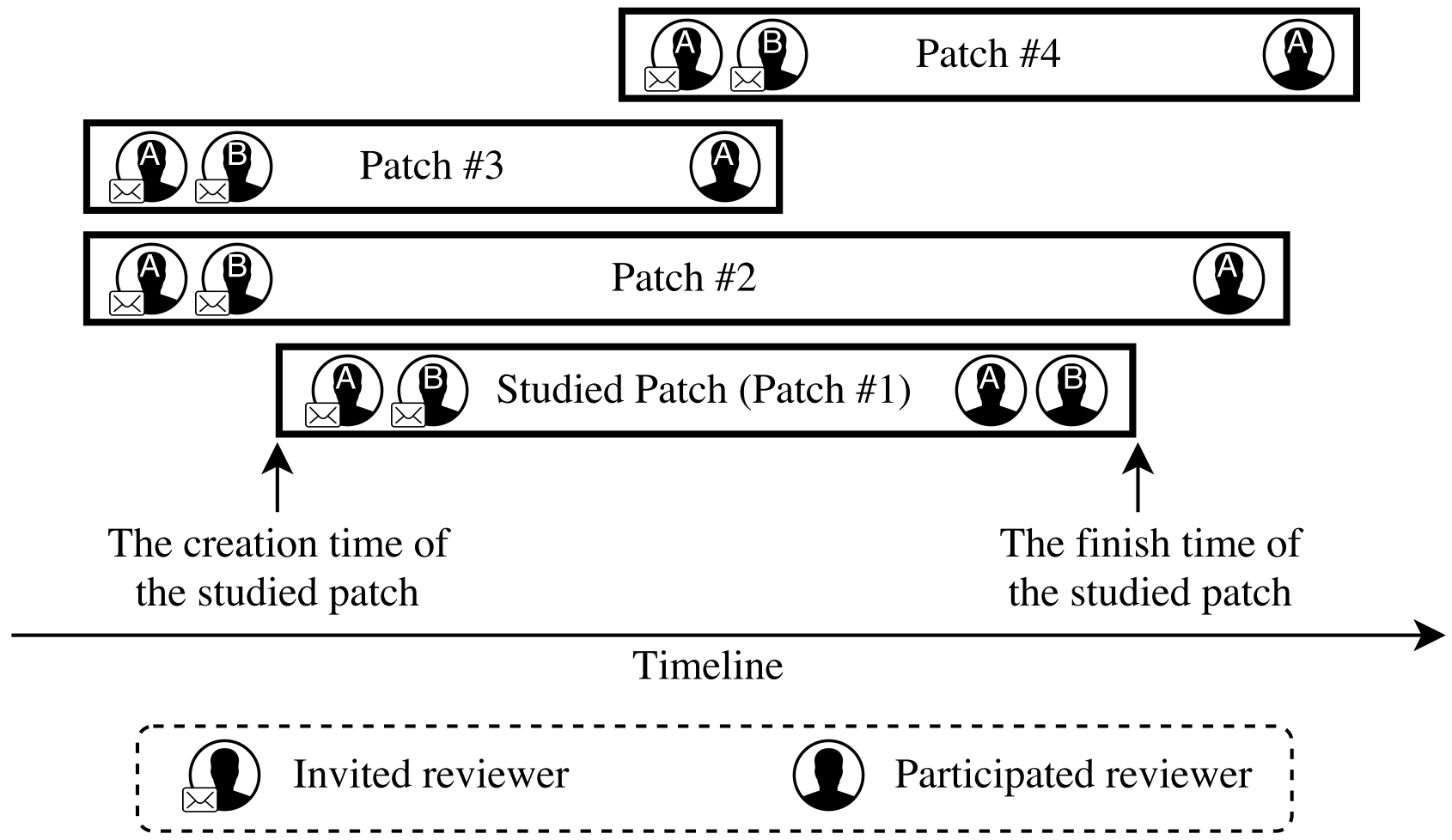

Fig. 4: An example of how to count the number of concurrent reviews and the number of remaining reviews.

Social Interaction Social interaction measures past activities that an invited reviewer had involved with patch authors or prior patches. *Familiarity between the Invited Reviewer and the Patch Author* counts the number of prior patches that an invited reviewer had reviewed for the patch author. *Median Number of Comments* measures a median number of messages that an invited reviewer had posted on prior patches that impact the same subsystem as the studied patch. *Review Participation Rate* measures a proportion of prior patches that impact the same subsystem as the studied patch and an invited reviewer responded to the review invitation. More specifically, we measure the review participation rate of an invited as described below:

$$\text{Review Participation Rate} = \frac{\#\text{Responded review invitations}}{\#\text{Received review invitations}} \tag{1}$$

*Number of Received Review Invitations* counts the number of prior patches that an invited reviewer had received a review invitation. *Core Member Status* is identified based on the permission to approve or abandon a patch (i.e., providing a review score of +2 or -2).[13] The core member status is assigned as TRUE if a reviewer has provided a review score of +2 or -2 to prior patches, FALSE otherwise. In this work, we only consider an approver role as a core member. A reviewer with a verifier role may also has a core member status, however this verifying task is often performed by automatic tools (e.g., Continuous Integration tools) (McIntosh et al, 2014).

[13] https://gerrit-review.googlesource.com/Documentation/access-control.html#examples_developer

***Reviewer Experience Dimension*** Reviewer experience dimension measures the related experience that an invited reviewer has on a patch. *Reviewer Code Authoring Experience* measures how many prior patches that an invited reviewer had authored. To measure the code authoring experience, we first measure code authoring experience for each module that is impacted by the studied patch using a calculation of $\frac{a(D,M)}{C(M)}$ (Bird et al, 2011), where $a(D,M)$ is the number of prior patches that the invited reviewer $D$ had made to module $M$. $C(M)$ is the total number of prior patches that were made to $M$. Then, we calculate an average of the code authoring experience of these impacted modules. *Reviewer Reviewing Experience* measures how many prior patches that an invited reviewer had reviewed. To measure the reviewing experience, we first measure reviewing experience for each module that is impacted by the studied patch using a calculation of $\frac{\sum_{k=1}^{r(D,M)} \frac{1}{R(k)}}{C(M)}$ (Thongtanunam et al, 2016b), where $r(D,M)$ is the number of prior patches made to module $M$ which the invited reviewer $D$ had reviewed. $R(k)$ is the total number of reviewers who reviewed patch $k$. $C(M)$ is the total number of prior patches that were made to $M$. Finally, we calculate an average of the reviewing experience of the studied in these impacted modules.

***Patch Characteristics Dimension*** Patch characteristic dimension measures characteristics of the studied patch. *Patch Size* counts how many lines of code that were changed in the studied patch. *Patch Author Code Authoring Experience* measures how many prior patches that the patch author had authored. To measure the code authoring experience, we use the same calculation as we use for the reviewer code authoring experience. *Patch Author Reviewing Experience* measures how many prior patches that the patch author has reviewed. To measure the reviewing experience, we use the same calculation as we use for the reviewer reviewing experience.

## 4.3 Model Construction (MC)

We construct nonlinear logistic regression models to determine the likelihood that an invited reviewer will participate in a review. The nonlinear logistic regression model is a logistic regression model that provides more flexible curve-fitting methods than a linear logistic regression model. We adopt the model construction approach of Harrell Jr. (2002), which enables a more accurate and robust fit of the dataset than the linear logistic regression model construction approach, while carefully considering the potential for overfitting. We use the studied metrics as independent variables. The dependent variable is assigned as the value of TRUE if an invited reviewer responded to the review invitation, and FALSE otherwise. Figure 2 provides an overview of our model construction approach which consists of three main steps. We describe each step below.

**(MC1) Remove highly-correlated and redundant independent variables.** Using highly correlated or redundant independent variables in regression models can create distorted and exaggerated relationships between the independent variables and the dependent variable, which lead to spurious conclusions (Mason and Perreault Jr, 1991; Tantithamthavorn et al, 2016). To analyze the correlation between the independent variables, we perform Spearman rank correlation tests ($\rho$) (Spearman, 1904). Then, we construct a hierarchical overview of the correla-tion using the variable clustering analysis technique (Sarle, 1990). For a cluster of highly correlated variables, we select only one variable as a representative variable for that cluster. Suggested by Hinkle et al (1998), Spearman correlation coefficient values ($\rho$) greater than 0.7 are considered as strong correlations. Therefore, we use a threshold of $|\rho| > 0.7$, which is also used in prior studies (McIntosh et al, 2016; Thongtanunam et al, 2016a). We repeat this process until the Spearman correlation coefficient values of all clusters are less than 0.7.

After the correlation analysis, we also perform a redundancy analysis to check whether the surviving variables provide a unique signal or not (Harrell Jr., 2015b, p. 80). We use the `redun` function in the `Hmisc` package (Harrell Jr., 2015a) to detect redundant variables and remove them from our models.

**(MC2) Construct nonlinear logistic regression models.** To construct a nonlinear logistic regression model with a low risk of overfitting, we need to consider degrees of freedom that can be allocated to the model. A model that uses degrees of freedom more than a dataset can support can be overfit to that dataset (Harrell Jr., 2002). Therefore, we estimate a budget for degrees of freedom using a calculation of $\frac{min(T,F)}{15}$ (Harrell Jr., 2002), where $T$ is the number of instances where the dependent variable is TRUE, and F is the number of instances where the dependent variable is FALSE. The total allocated degrees of freedom should never exceed the budgeted degrees of freedom. Once we have budgeted degrees of freedom, we allocate the degrees of freedom to the independent variables that survive from our correlation and redundancy analysis.

To model nonlinear relationships, we allocate degrees of freedom to an independent variable. Similar to prior work (McIntosh et al, 2016), the degree of freedom is allocated based on Spearman multiple $\rho^2$ value, which indicates the potential of sharing a nonlinear relationship between an independent and a dependent variable. The larger the Spearman multiple $\rho^2$ value is, the higher potential of sharing a nonlinear relationship. We use the `spearman2` function in the `rms` R package (Harrell Jr., 2015c) to calculate the Spearman multiple $\rho^2$ value for each independent variable. Then, we manually identify a group of independent variables based on their Spearman multiple $\rho^2$ values. Although there may be a large number of budgeted degrees of freedom, we only allocate three to five degrees of freedom to a group with the high Spearman multiple $\rho^2$ value and allocate one degree of freedom (i.e., a linear fit) to a group with a low Spearman multiple $\rho^2$ values. This is because allocating too many degrees of freedom may lead a model to overfit, which will exaggerate spurious relationships between an independent variable and the dependent variable (McIntosh et al, 2016).

Once we have removed highly-correlated and redundant variables and allocated degrees of freedom to the surviving variables, we construct a nonlinear logistic regression model. Similar to prior work (McIntosh et al, 2016; Thongtanunam et al, 2016a), we use restricted cubic splines (also called natural splines) to fit the data using the allocated degrees of freedom. We use the restricted cubic splines because the smooth characteristic of cubic splines fits highly curved functions better than a linear splines (Harrell Jr., 2002, p. 20). In addition, the restricted cubic splines also behave better than the unrestricted cubic splines in the tails of functions, i.e., before the first change in direction of functions and after the last change in direction of functions.

### 4.4 Model Analysis (MA)

We analyze nonlinear logistic regression models to determine the performance of the models, and to quantitatively understand the relationship between the independent variables and the participation decision of an invited reviewer. Figure 2 provides an overview of our model analysis approach which consists of three main steps. We describe each step below.

**(MA1) Evaluate the models.** We evaluate the performance of our models using Area Under the receiver operating Curve (AUC) (Hanley and McNeil, 1982) and Brier score (Brier, 1950). AUC and Brier score are threshold-independent measures, i.e., the measurement does not rely on the probability threshold (e.g., 0.5) (Tantithamthavorn and Hassan, 2018). Moreover, AUC and Brier score are robust to the data where the distribution of a dependent variable is skewed (Fawcett, 2006). Nonetheless, we also measure precision, recall, and F-measure which are commonly used in software engineering literature (Elish and Elish, 2008; Foo et al, 2015; Tantithamthavorn et al, 2015; Zimmermann et al, 2005). Below, we describe each of the performance measures:

- **AUC value** measures how well a model can discriminate between two groups of the dependent variables (i.e., a reviewer who will respond to a review invitation and who will not). An AUC value of 1 indicates a perfect discrimination ability while an AUC value of 0.5 indicates that the discrimination ability of the model is not better than random guessing.
- **Brier score** measures the error of the predicted probability of the model. A Brier score of 0.25 indicates that the accuracy of the model is not better than random guessing. The lower the Brier score is, the less error the predicted probability of the model is. The Brier score is calculated as described below:

$$\text{Brier score} = \frac{1}{N}\sum_{i=1}^{N}(\text{predicted probability}_i - \text{actual outcome}_i)^2 \qquad (2)$$

  where $N$ is the total number of instances. The actual outcome is 1 if the dependent variable of instance $i$ is TRUE (i.e., an invited reviewer responded to the review invitation) and 0 otherwise.
- **Precision** measures the correctness of a prediction model in predicting whether invited reviewers will respond to the review invitation. More specifically, precision is a ratio of the correctly predicted instances where our models predict that invited reviewers will respond to the review invitation.
- **Recall** measures the completeness of a prediction model in predicting whether invited reviewers will respond to the review invitation. More specifically, recall is a ratio of the correctly predicted instances that invited reviewers had responded to the review invitation.
- **F-measure** is the weighted harmonic mean of precision and recall. It is a measure that indicates the balance between precision and recall. The F-measure value ranges from 0 to 1. The higher the F-measure value is, the better overall prediction performance of the model is. F-measure is calculated as described below:

$$\text{F-measure} = 2 \times \frac{\text{precision} \times \text{recall}}{\text{precision} + \text{recall}} \qquad (3)$$

To validate our results, we use the out-of-sample bootstrap validation technique (Efron, 1983). The key intuition of the out-of-sample bootstrap is that the relationship between the studied dataset and the theoretical population from which it is derived is asymptotically equivalent to the relationship between the bootstrap samples and the studied dataset (Efron, 1983). Tantithamthavorn et al (2017b) also find that for logistic regression models, the out-of-sample bootstrap validation approach is the least biased model validation technique (i.e., has the least difference between the performance estimates and the model performance on unseen data) compared to other commonly-used validation techniques like K-fold cross validation. Moreover, the out-of-sample bootstrap validation approach also produces more stable performance estimates (i.e., there is little change in performance estimates when repeating the experiments).

The out-of-sample bootstrap consists of three steps. First, we randomly draw a bootstrap sample with replacement from the original dataset. A bootstrap sample has the same size as the original dataset. Then, we construct a prediction model using the bootstrap sample. Finally, we test and measure the performance of the bootstrap model using the instances that are not in the bootstrap sample. We repeat this process with 1,000 iterations and compute an average of each performance measure.

**(MA2) Estimate the power of explanatory.** To identify the most influential factors that can have on the participation decision of an invited reviewer, we estimate the explanatory power that each independent variable can contribute to the fit of the model. To do so, similar to prior work (McIntosh et al, 2016; Thongtanunam et al, 2016a), we measure the Wald statistics (Wald $\chi^2$) and its statistical significance ($p$-value) using the `anova` function in the `rms` R package (Harrell Jr., 2015c). Since the independent variables that are allocated more than one degrees of freedom, are represented with several model terms, we use the Wald statistics to jointly test a set of model terms for each independent variable. The larger the Wald $\chi^2$ of a variable is, the larger the explanatory power that the variable contributes to the model.

In addition, to determine ranks of the explanatory power of the variables, we use the Scott-Knott Effect Size Difference (ESD) test (Tantithamthavorn et al, 2017b). The Scott-Knott ESD test is an enhanced variant of the Scott-Knott test. The Scott-Knott ESD test is more appropriate for our datasets since it will mitigate the skewness of an input dataset and merge any two statistically distinct groups that have a negligible effect size into one group. To do so, we construct our model using a bootstrap sample (i.e., a dataset that is randomly sampled with replacement from the original dataset). Then, we estimate Wald $\chi^2$ of the bootstrap model. We repeat this process for 1,000 bootstrap samples. Finally, we use the `sk_esd` function in the `ScottKnottESD` R package (Tantithamthavorn et al, 2017b) to cluster the distribution of Wald $\chi^2$ of the 1,000 bootstrap models into statistically distinct ranks.

**(MA3) Examine the relationships between the variables and the participation decision.** The power of explanatory and the Scott-Knott ESD test provide (1) the magnitudes of the impact of each independent variable on model performance; and (2) the rank of the power of explanatory of the independent variables. However, they do not provide direction or shape of the relationship between independent variable and the likelihood that an invited reviewer will participate in a review. To further observe the direction of the relationship between the indepen-dent variable and the likelihood, we use the `Predict` function in the `rms` R package (Harrell Jr., 2015c) to plot the likelihood against the particular variable, while controlling for the other variables at their median values.

In addition, to quantify the impact that each independent variable can have on the likelihood that an invited reviewer will participate in a review, we estimate the partial effect of the independent variables using odds ratio (Harrell Jr., 2002, p. 220). Odds ratio indicates the change to the likelihood when there is a change in the value of the independent variable. A positive odds ratio indicates an increasing relationship while a negative odds ratio indicates a decreasing relationship. A large odds ratio indicates a large partial effect that the independent variable has on the likelihood.

## 5 Case Study Results

In this section, we present the results of our case study according to our three research questions.

### (RQ1) How often do patches suffer from the unresponded review invitations?

*Approach*

To address the RQ1, we analyze descriptive statistics of the number of reviewers who did not respond to the review invitation of patches. In particular, we count how many patches that have reviewers who did not respond to the review invitation. Furthermore, we investigate whether or not inviting many reviewers can decrease a chance of having an invited reviewer who did not respond to the review invitation. To do so, we first plot the number of invited reviewers against the number of reviewers who did not respond to the review invitation in a patch using hexbin plots (Carr et al, 1987). We also use Kendall rank correlation coefficient ($\tau$) to measure the correlation between the number of invited reviewers and the number of reviewers who did not respond to the review invitation. Instead of using the commonly-used Spearman rank correlation (Spearman, 1904), we use Kendall rank correlation in order to provide a more robust and more interpretable correlation (Croux and Dehon, 2010; Newson, 2002). The Kendall's $\tau$ is considered as trivial for $|\tau| < 0.1$, small for $0.1 \leq |\tau| < 0.3$, medium for $0.3 \leq |\tau| < 0.5$, and large for $0.5 \leq |\tau| \leq 1$ (Cohen, 1992). A positive value of Kendall's $\tau$ indicates an increasing relationship between the variables while a negative value indicates a decreasing relationship between the variables. A value of zero indicates an absence of a relationship.

*Results*

**Observation 1 — 16%-66% of the patches have at least one invited reviewer who did not respond to the review invitation.** We find that 24,367 of 36,771 (66%) patches in the Android dataset, 3,039 of 18,716 (16%) patches in the LibreOffice dataset, 24,589 of 108,788 (23%) patches in the OpenStack dataset, and

30,630 of 65,815 (47%) patches in the Qt dataset are not responded to by at least one invited reviewer. These results suggest that when a patch author invites reviewers for a new patch, there is a high chance (especially in the Android and Qt datasets) that one of the invited reviewers will not respond to the review invitation. Moreover, 1,343 of 3,039 (44%) patches in the LibreOffice dataset that were not responded to by an invited reviewer do not have any invited reviewers participated in. We also observe 19%, 4% and 5% of such patches in the Android, OpenStack and Qt datasets.

We also find that on average, there are one (LibreOffice, Qt and OpenStack datasets) to two (Android dataset) invited reviewers who did not respond to the review invitation. Figure 5 shows the distributions of reviewers who did not respond to the review invitation in a patch using a violin plot. We use a violin plot to summarize the distributions in vertical curves. Since the number of reviewers can vary among patches, we analyze the proportion of reviewers who did not respond to the review invitation in a patch instead of the actual number of reviewers. The wider the violin plot is, the more patches that have the corresponding proportion of reviewers who did not respond to the review invitation are. We observe that at the median, 33%(OpenStack)-67%(Android) of the invited reviewers did not respond to the review invitation. We also observe that at the median of the Android dataset (i.e., where patches have 67% of invited reviewers who did not respond to the review invitation), patch authors often invited three reviewers and two of them did not respond to the review invitation. At the median of the LibreOffice dataset, patch authors often invited two reviewers and one of them did not respond to the review invitation. At the median of the OpenStack dataset, patch authors often invited three reviewers and one of them did not respond to the review invitation. At the median of the Qt dataset, patch authors often invited two reviewers and one of them did not respond to the review invitation.

**Observation 2 — The number of invited reviewers shares an increasing relationship with the number of reviewers who did not respond to the review invitation.** Figure 6 shows the number of invited reviewers against the number of reviewers who did not respond to the review invitation using hexbin plots (Carr et al, 1987). Hexbin plots are scatter plots that represent several data points with hexagon-shaped bins. The darker the shade of the hexagon, the more data points that fall within the bin. We also draw a line with 95% confidence

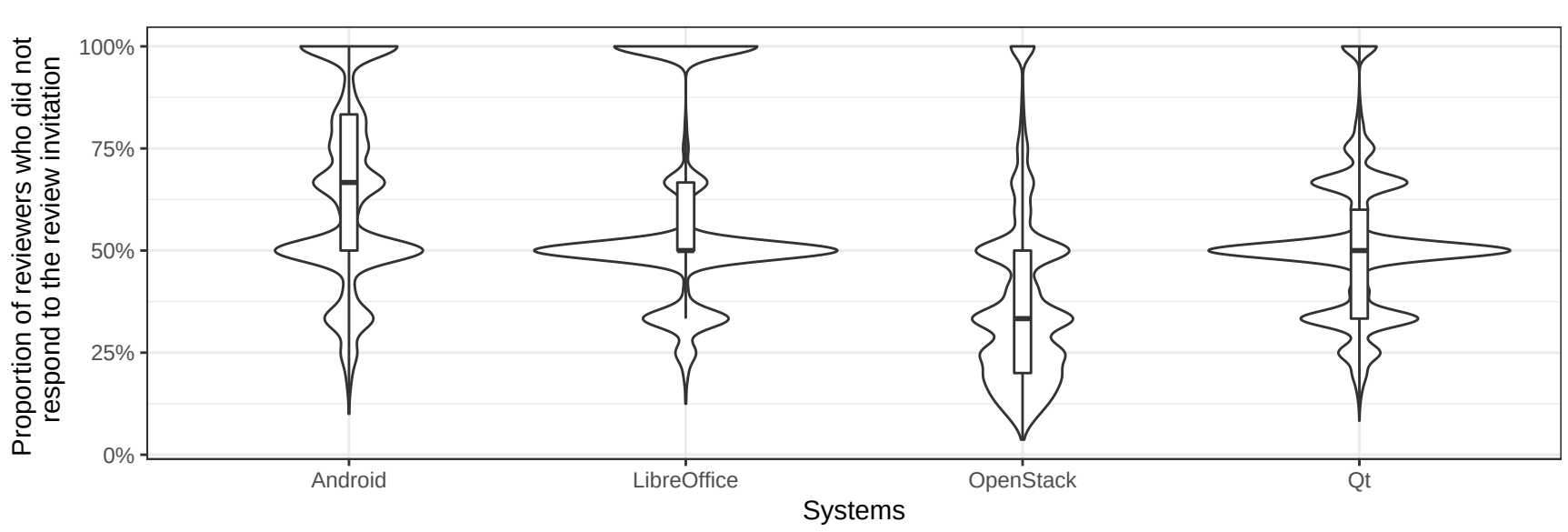

Fig. 5: The distributions of proportion of reviewers who did not respond to the review invitation in a patch.

interval (the gray area) over the plot in order to observe the correlation. We only observe the plots where the gray area is narrow, since a wide gray area indicates the lack of data points. Figure 6 shows that the more invited reviewers the patch has, the more reviewers who did not respond to the patch. We also find that the Kendall rank correlation coefficient $\tau$ is +0.719 (large) in the Android dataset, +0.544 (large) in the LibreOffice dataset, +0.344 (medium) in the OpenStack dataset, and +0.671 (large) in the Qt dataset.

*A large number of patches (i.e., 16%-66%) have at least one invited reviewer who did not respond to the review invitation. Moreover, the results suggest that if patch authors invite more reviewers, the chance of having a non-responding reviewer tends to increase.*
*(Observations 1-2).*

**(RQ2) Can human factors help determining the likelihood of the participation decision of reviewers?**

The results of RQ1 suggest that if patch authors invite more reviewers, the chance of having a non-responding reviewer tends to increase. In other words, the approach of inviting more reviewers to increase review participation is becoming less efficient as the number of invited reviewers is increasing. Therefore, instead of inviting more reviewers, a better understanding of factors playing a role in this process can be of help to patch authors. To better understand the factors that can have an impact on the participation decision of reviewers, we construct a prediction model to determine the likelihood of the participation decision of reviewers (i.e., whether an invited reviewer will participate in a review). Unlike the prior studies that mainly use reviewer experience and patch characteristics (Balachandran, 2013; Thongtanunam et al, 2015b; Xia et al, 2015; Yu et al, 2014; Zanjani et al, 2016), we use human factors, reviewer experience, and patch characteristics to construct our prediction models. Below, we describe our approach to address our RQ2, then present our results and observations.

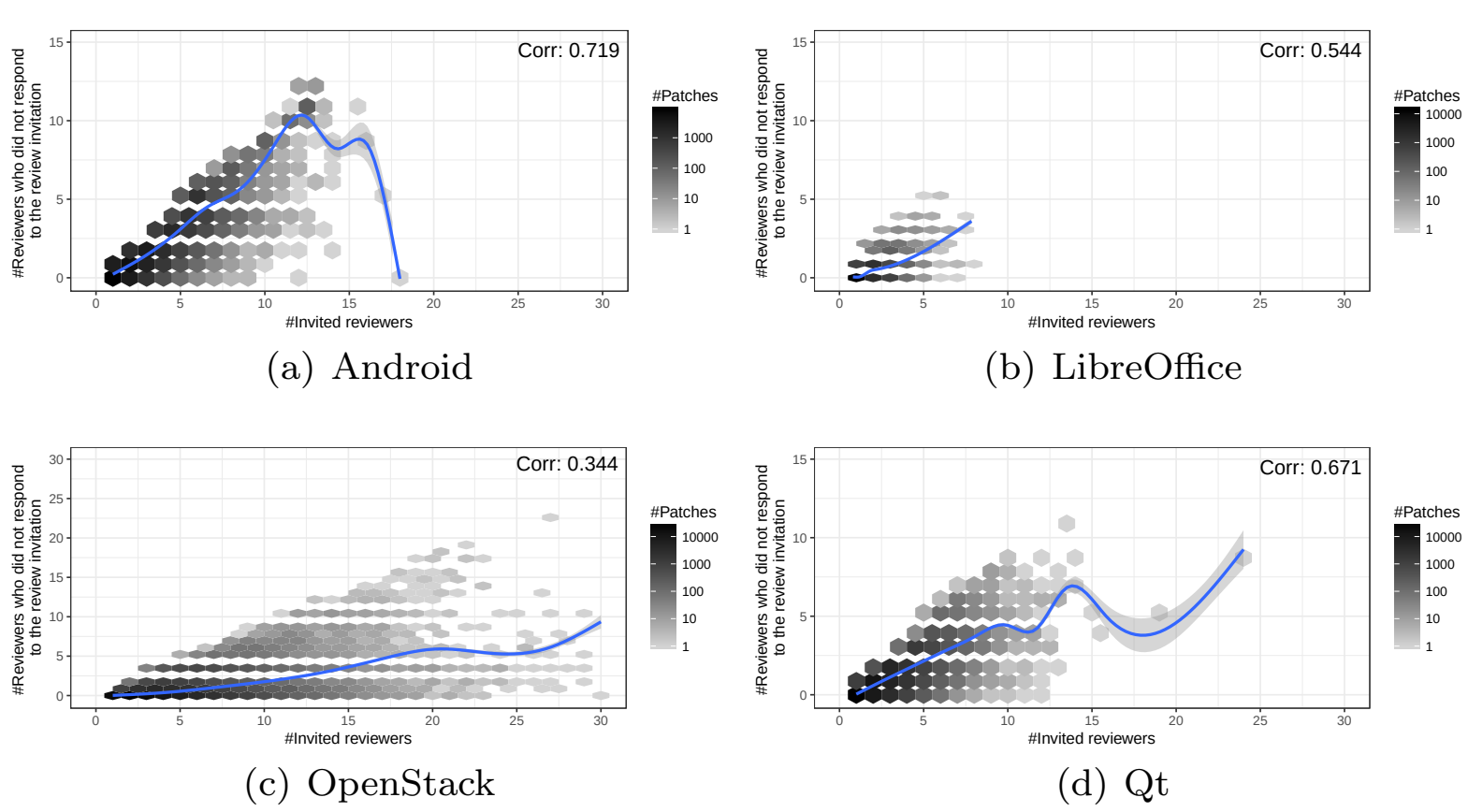

(a) Android (b) LibreOffice

(c) OpenStack (d) Qt

Fig. 6: The hexbin plots of the number of invited reviewers against the number of reviewers who did not respond to the review invitation. The gray area shows the 95% confidence interval.

Table 3: Summary of instances of the studied datasets.

| | #TRUE Instances | #FALSE Instances |
|---|---|---|
| Android | 77,720 (59%) | 54,411 (41%) |
| LibreOffice | 25,905 (88%) | 3,572 (12%) |
| OpenStack | 421,927 (90%) | 44,593 (10%) |
| Qt | 155,367 (77%) | 47,196 (23%) |

*Approach*

To address our RQ2, for each dataset, we construct two nonlinear logistic regression models to predict whether or not an invited reviewer will participate in the review. One is our proposed model that includes human factors and another model is the baseline model that does not include human factors. In particular, we use all metrics listed in Table 2 as independent variables for our proposed model. For the baseline model, we use only metrics in the reviewer experience and patch characteristics dimensions as independent variables. For each patch, both proposed and baseline models predict the outcome for each invited reviewer, i.e., TRUE if an invited reviewer responded to the review invitation, and FALSE otherwise. In total, we use the datasets of 132,131 (Android), 29,477 (LibreOffice), 466,520 (OpenStack), and 202,563 (Qt) instances for this RQ. Table 3 shows the number of TRUE and FALSE instances of each studied dataset. Then, we construct our prediction models according to our model construction approach (see Section 4.3), which we discuss each step below.

**(MC1) Remove highly-correlated and redundant independent variables.** We remove independent variables that are highly correlated with one another based on the variable clustering analysis. For example, Figure 7 shows the hierarchical clustering of variables of the OpenStack dataset. We find that the number of remaining reviews and the number of concurrent reviews are highly correlated, i.e., a Spearman's $|\rho|$ value is greater than 0.7. Therefore, we choose the number of concurrent reviews as a representative variable of this cluster because the number of concurrent reviews is a more straightforward metric of review workload than the number of remaining reviews (i.e., a large number of concurrent reviews indicate that the invited reviewer has been already involved in many patches and may not be able to review the new patch). We perform the variable clustering analysis again, and we find that none of the surviving variables are highly correlated. We perform a similar process for Android and Qt datasets. However, none of the independent variables in the Android, LibreOffice and Qt datasets have a Spearman's $|\rho|$ value greater than 0.7. Hence, we use all of the metrics for the Android, LibreOffice and Qt datasets.

We also check for the redundant variables. However, we find that there are no surviving variables that have a fit with an $R^2$ greater than 0.9 for all four datasets. Hence, we use all the surviving independent variables to construct the models.

**(MC2) Construct nonlinear logistic regression model.** Before we construct a model, we estimate a budget for degrees of freedom. Table 4 shows the budgeted degrees of freedom for each dataset. Then, we allocate degrees of freedom to the independent variables. For example, Figure 8 shows Spearman multiple $\rho^2$ of each independent variable for the Android dataset. Figure 8 shows that the review participation rate has the largest Spearman multiple $\rho^2$ value. The reviewer code authoring experience, the median number of comments, the reviewer reviewing experience, and the number of concurrent reviews have medium multiple $\rho^2$ value. The other variables have small Spearman multiple $\rho^2$ values. Thus, we allocate three degrees of freedom to the review participation rate and the variables that have medium multiple $\rho^2$ value. We allocate one degree of freedom to the other variables. We perform the similar process for the OpenStack and Qt datasets. Table 4 shows the number of degrees of freedom that we allocate to each independent variable. Finally, we construct the models using the allocated degrees of freedom. The final allocated degrees of freedom are decided by the restricted cubic splines according to the compatibility of the independent variable values. For example, although we allocate three degrees of freedom to a variable, it is possible that the restricted cubic splines will allocate only one degree of freedom to the variable if the variable values cannot support to have a change in the direction of its function.

*Results*

**Observation 3 — Our proposed models that include human factors achieve an AUC value of 0.82-0.89, a Brier score of 0.06-0.13, a precision of 0.68-0.78, a recall of 0.24-0.73, and an F-measure of 0.35-0.75.** Table 5 shows an average of performance measures of our proposed models. Our proposed models achieve an average AUC value of 0.82(Qt)-0.89(Android) and a Brier score of

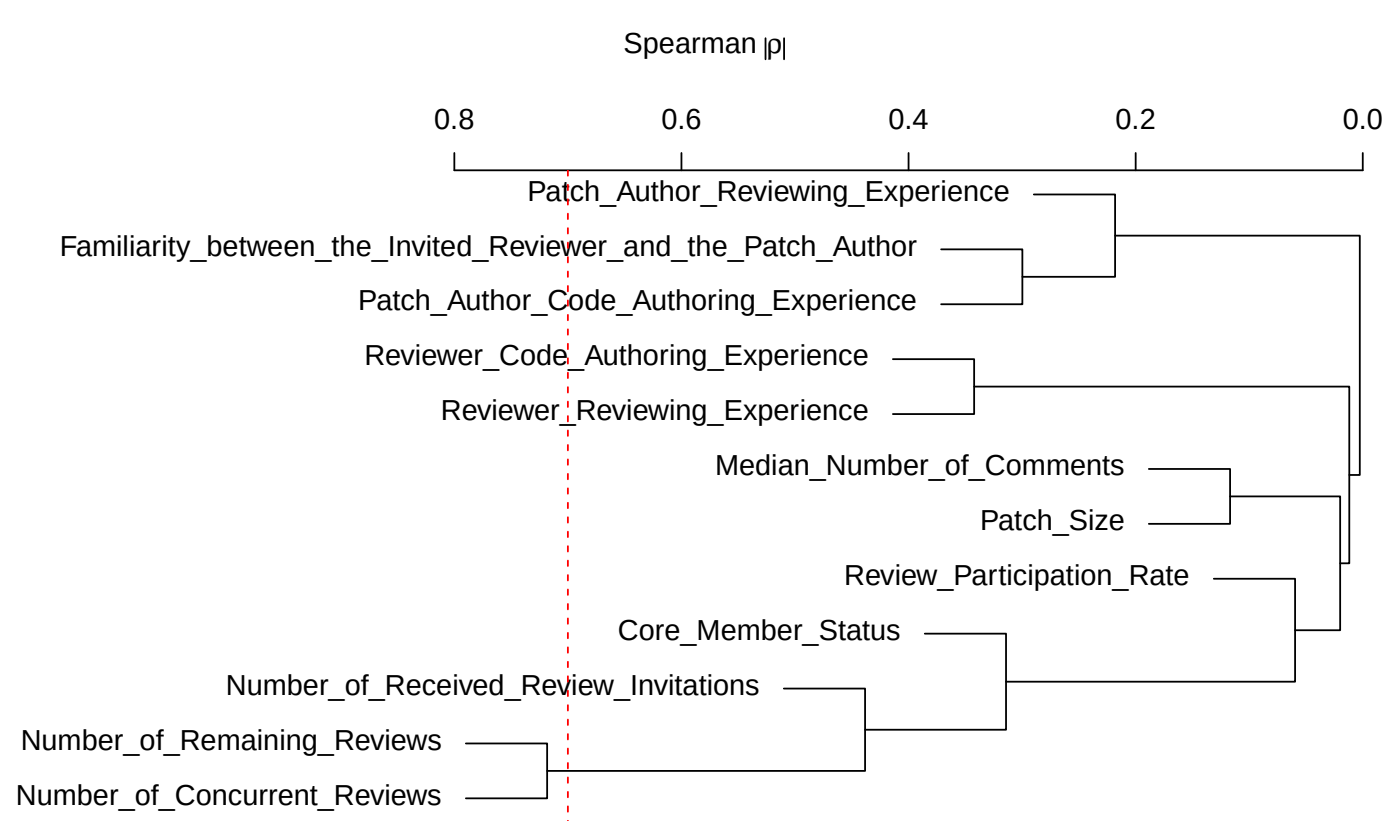

Fig. 7: Hierarchical clustering of independent variables of the OpenStack dataset.

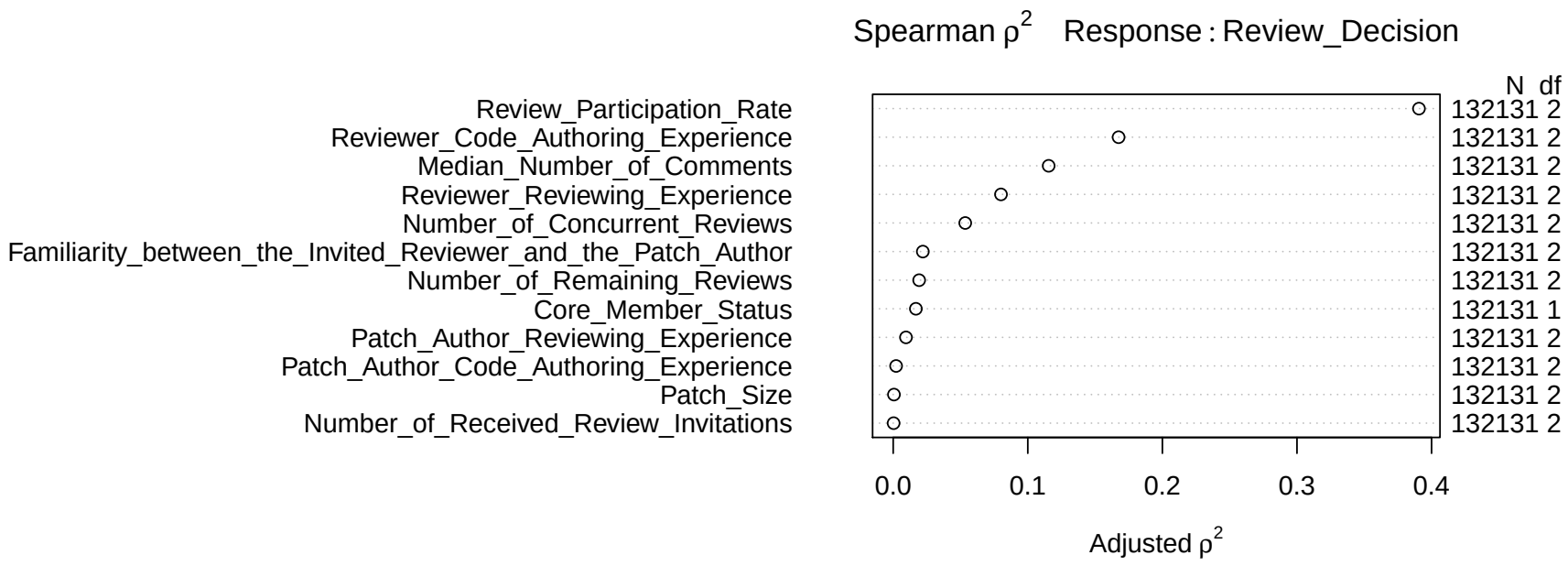

Fig. 8: Dotplot of the Spearman multiple $\rho^2$ of each independent variable of the Android dataset. The higher the Spearman multiple $\rho^2$ is, the more potential the variable has.

Table 4: Summary of the degrees of freedom. The metrics in the human factors dimension are only included in our proposed models.

| | **Android** | | **LibreOffice** | | **OpenStack** | | **Qt** | |
|---|---|---|---|---|---|---|---|---|
| Budgeted Degrees of Freedom | 3,627 | | 238 | | 2,972 | | 3,146 | |
| Spent Degrees of Freedom | 17 | | 17 | | 12 | | 15 | |
| | Overall | Nonlinear | Overall | Nonlinear | Overall | Nonlinear | Overall | Nonlinear |
| ***Human Factors Dimension*** | | | | | | | | |
| Number of Concurrent Reviews | 2 | 1 | 2 | 1 | 1 | — | 1 | — |
| Number of Remaining Reviews | 1 | — | 3 | 2 | † | | 1 | — |
| Familiarity between Invited Reviewer and Patch Author | 1 | — | 1 | — | 1 | — | 2 | 1 |
| Median Number of Comments | 2 | 1 | 1 | — | 1 | — | 1 | — |
| Review Participation Rate | 2 | 1 | 2 | 1 | 2 | 1 | 2 | 1 |
| Number of Received Review Invitations | 1 | — | 1 | — | 1 | — | 1 | — |
| Core Member Status | 1 | — | 1 | — | 1 | — | 1 | — |
| ***Review Experience Dimension*** | | | | | | | | |
| Reviewer Code Authoring Experience | 2 | 1 | 1 | — | 1 | — | 2 | 1 |
| Reviewer Reviewing Experience | 2 | 1 | 2 | 1 | 1 | — | 1 | — |
| ***Patch Characteristics Dimension*** | | | | | | | | |
| Patch Size | 1 | — | 1 | — | 1 | — | 1 | — |
| Patch Author Code Authoring Experience | 1 | — | 1 | — | 1 | — | 1 | — |
| Patch Author Reviewing Experience | 1 | — | 1 | — | 1 | — | 1 | — |

†: This variable is removed during the variable clustering analysis.
—: Nonlinear degrees of freedom are not allocated.

Table 5: The estimated performance measures of our proposed models and the baseline models with the standard deviation of the performance distributions of the models that are trained using 1,000 bootstrap samples. The *%Improvement* row shows the percentage improvement of our proposed model compared to the baseline model.

| Measure | Android | | LibreOffice | | OpenStack | | Qt | |
|---|---|---|---|---|---|---|---|---|
| | Proposed | Baseline | Proposed | Baseline | Proposed | Baseline | Proposed | Baseline |
| AUC value | 0.89±0.001 | 0.77±0.002 | 0.86±0.004 | 0.78±0.005 | 0.86±0.001 | 0.62±0.002 | 0.82±0.001 | 0.70±0.002 |
| %Improvement | +16% | | +10% | | +39% | | +17% | |
| Brier score | 0.13±0.001 | 0.19±0.001 | 0.08±0.001 | 0.09±0.002 | 0.06±0.000 | 0.09±0.000 | 0.13±0.001 | 0.16±0.001 |
| %Improvement | -32% | | -11% | | -33% | | -19% | |
| Precision | 0.78±0.003 | 0.61±0.003 | 0.70±0.022 | 0.77±0.038 | 0.73±0.005 | 0.75±0.037 | 0.68±0.005 | 0.60±0.027 |
| %Improvement | +28% | | -9% | | -3% | | +13% | |
| Recall | 0.73±0.003 | 0.67±0.009 | 0.24±0.012 | 0.07±0.005 | 0.25±0.003 | 0.01±0.001 | 0.35±0.004 | 0.01±0.001 |
| %Improvement | +9% | | +243% | | +2,400% | | +3,400% | |
| F-measure | 0.75±0.002 | 0.64±0.004 | 0.35±0.013 | 0.13±0.009 | 0.38±0.003 | 0.02±0.001 | 0.46±0.003 | 0.03±0.002 |
| %Improvement | +17% | | +169% | | +1,800% | | +1,433% | |

0.06(OpenStack)-0.13(Android and Qt). Moreover, our proposed models achieve a precision of 0.68(Qt)-0.78(Android), a recall of 0.24(LibreOffice)-0.73(Android), and an F-measure of 0.35(LibreOffice)-0.75(Android). The LibreOffice, OpenStack and Qt models achieve a relatively low recall (i.e., 0.24, 0.25 and 0.35 respectively), indicating that the completeness of the identification of our proposed models for the LibreOffice, OpenStack and Qt datasets is relatively low. One possible explanation is the imbalanced data. We observe that the Android dataset tends to have the balanced number of instances with TRUE (59%) and FALSE (41%). On the other hand, the LibreOffice, Qt and OpenStack datasets tend to have the imbalanced number of instances (i.e., the majority of instances are 88%, 90% and 77% for the LibreOffice, OpenStack and Qt datasets, respectively). We further discuss the improvement of recall of our prediction models in Section 8.1

**Observation 4 — AUC increases by 10%-39%, Brier score decreases by 11%-33%, and F-measure increases by 17%-1,800% when we include human factors into the models** Table 5 shows that the baseline models (i.e., the models that do not include human factors) achieve an average AUC value of 0.62(OpenStack)-0.78(LibreOffice), a Brier score of 0.09(LibreOffice and Qt)-0.19(Android), a precision of 0.60(Qt)-0.77(LibreOffice), a recall of 0.01(OpenStack and Qt)-0.67(Android), and an F-measure of 0.02(OpenStack)-0.64(Android). We measure the performance improvement between our proposed models and the baseline models using a calculation of $\frac{P_{proposed} - P_{baseline}}{P_{baseline}}$, where $P$ is the value of performance measures. Table 5 shows the performance improvement for all of our five performance measures. We observe that AUC increases by 10%-39%, Brier score decreases by 11%-33%, and F-measure increases by 17%-1,800%. These results indicate that including human factors into the prediction models of review participation decision increases the performance.

Although we find a small decrease in precision for the LibreOffice and OpenStack datasets (9% and 3%, respectively), we find a large increase in recall (243% and 2,400%, respectively). The F-measure (which combines precision and recall measures) shows that our proposed models outperform the baseline models with an increase of 169% for the LibreOffice dataset and 1,800% for the OpenStack dataset.

In addition, we observe that the baseline models perform poorly for the LibreOffice, OpenStack and Qt datasets. Therefore, we further investigate this result by quantifying the difference between the distributions of each independent variable when the dependent variable is TRUE and FALSE using Cliff's Delta (Cliff, 1993, 1996). A negligible difference between the distributions of the independent variable indicates that the variable does not provide a unique signal to the prediction model. We find that, in the LibreOffice dataset, all variables of the patch characteristics dimension have a negligible differences while the variables of the reviewer experience dimension have negligible to medium differences. On the other hand, the variables of the human factors dimension such as review participation rate have small to large differences between their distributions. We also find similar results in the OpenStack and Qt datasets.

> *The results suggest that including human factors into the prediction models of review participation decision increases the performance. Thus, leading to a more accurate description of this process.*

**(RQ3) What are the factors mostly associated with participation decision?**

Our RQ2 results show that including human factors into a prediction model can increase the ability to determine the likelihood of the participation decision. Hence, in this RQ, we quantitatively understand the relationship between each factor in our proposed models and the likelihood that an invited reviewer will participate in a review. Below, we describe our approach to address our RQ3, then present our results and observations.

*Approach*

To address our RQ3, we analyze our prediction models according to our model analysis approach (see Section 4.4).

*Results*

Table 6 shows the explanatory power of each independent variable. The *Overall* column shows the proportion of the Wald $\chi^2$ of the entire model fit that is attributed to that independent variable. The *Nonlinear* column shows the proportion of the Wald $\chi^2$ of the entire model fit that is attributed to the nonlinear component of that independent variable. The larger the proportion of the Wald $\chi^2$ is, the larger the explanatory power that a particular independent variable contributes to the model. Furthermore, Table 6 shows that 4 of 13 independent variables to which we allocated nonlinear degrees of freedom, receive a significant boost to the explanatory power from the nonlinear component. This result indicates that the nonlinear modeling improves the fit of our models and provides a more precise relationship between the independent variables and the likelihood that an invited reviewer will participate in a review. Thus, future research should consider this approach as well.

Table 6 also shows the estimated partial effect of each independent variable that can have on the likelihood that an invited reviewer will participate in a review. The *Odds Ratio* column shows the partial effect based on the shifted value of the variable that is shown in the *Shifted Value* column. The *Shifted Value* column shows an inter-quartile range of the variable values, which is used to estimate the partial effect shown in the *Odds Ratio* column. Odds ratio is the difference in the likelihood that an invited reviewer will participate in a review when the corresponding variable value shifts from the first quartile to the third quartile of the data. A positive odds ratio indicates that that independent variable has a positive impact on the likelihood, while a negative odds ratio indicates the negative impact.

**Observation 5 — Review participation rate shares an increasing nonlinear relationship with the likelihood that an invited reviewer will participate in a review.** Table 6 shows that the review participation rate of an invited reviewer accounts for the largest proportion of Wald $\chi^2$ in three of our four models, suggesting that the review participation rate is mostly associated with the participation decision in the three studied systems.

Figure 9(a) shows the direction of the relationship between the review participation rate of an invited reviewer and the likelihood that an invited reviewer will participate in a review in the Android model. We also observe the similar relationship in the LibreOffice, OpenStack and Qt models. Table 6 shows that when the review participation rate increases from 41% to 89% in the Android model, the likelihood increases by 1077%. Similarly, when the review participation rate increases from 87% to 96%, 91% to 100% and 74% to 92% in the LibreOffice, OpenStack and Qt models respectively, the likelihood increases by 134%, 8,537% and 165%. These results suggest that the active reviewers with high participation rate are more likely to respond to a new review invitation than the reviewers who have a lower participation rate. Practitioners may simply use the participation rate as

Table 6: The explanatory power of the independent variables, grouped into statistically distinct ranks by Scott-Knott ESD tests, and the partial effect that our independent variables have on the likelihood that a reviewer will participate in a review.

| Rank | Variable | Proportion of $\chi^2$ Overall | Proportion of $\chi^2$ Nonlinear | Shifted Value | Odds Ratio |
|---|---|---|---|---|---|
| **Android** | | | | | |
| 1 | Review Participation Rate | 73%* | 1%* | 41%→89% | 1,047%↑ |
| 2 | Reviewer Code Authoring Experience | 15%* | 0%* | 0→0.19 | 55%↑ |
| 3 | Patch Author Reviewing Experience | 5%* | — | 0→0.02 | -4%↓ |
| 4 | Reviewer Reviewing Experience | 2%* | 0%° | 0→0.002 | 1%↑ |
| 5 | Number of Remaining Reviews | 2%* | — | 2→16 | -10%↓ |
| 6 | Number of Received Review Invitations | 1%* | — | 11→348 | -11%↓ |
| 7 | Number of Concurrent Reviews | 0%* | 0%* | 3→21 | 20%↑ |
| 8 | Patch Author Code Authoring Experience | 0%* | — | 0.13→0.93 | -15%↓ |
| 9 | Core Member Status | 0%* | — | 0→1 | 13%↑ |
| 10 | Familiarity between the Invited Reviewer and the Patch Author | 0%* | — | 0→6 | -1%↓ |
| 11 | Median Number of Comments | 0%° | 0%° | 1→2 | 3%↑ |
| 12 | Patch Size | 0%° | — | 7→178 | 0% |
| **LibreOffice** | | | | | |
| 1 | Reviewer Reviewing Experience | 29%* | 2%* | 0→0.315 | 756%↑ |
| 2 | Review Participation Rate | 25%* | 5%* | 87%→96% | 134%↑ |
| 3 | Number of Remaining Reviews | 18%* | 3%* | 0→2 | -57%↓ |
| 4 | Patch Author Reviewing Experience | 11%* | — | 0→1 | -94%↓ |
| 5 | Number of Concurrent Reviews | 7%* | 2%* | 2→12 | 106%↑ |
| 6 | Reviewer Code Authoring Experience | 6%* | — | 0→0.05 | 9%↑ |
| 7 | Patch Author Code Authoring Experience | 2%* | — | 0.12→0.97 | -30%↓ |
| 8 | Number of Received Review Invitations | 1%* | — | 90→997 | -13%↓ |
| 9 | Familiarity between the Invited Reviewer and the Patch Author | 1%° | — | 0→11 | 8%↑ |
| 10 | Median Number of Comments | 0%° | — | 0→28 | 436%↑ |
| 11 | Core Member Status | 0%° | — | 0→1 | 11%↑ |
| 11 | Patch Size | 0%° | — | 12→230 | 0% |
| **OpenStack** | | | | | |
| 1 | Review Participation Rate | 93%* | 14%* | 91%→100% | 7,657%↑ |
| 2 | Reviewer Code Authoring Experience | 3%* | — | 0→0.08 | 17%↑ |
| 3 | Patch Author Reviewing Experience | 2%* | — | 0→1 | -90%↓ |
| 4 | Patch Author Code Authoring Experience | 1%* | — | 0.02→0.3 | -8%↓ |
| 5 | Reviewer Reviewing Experience | 0%* | — | 0→1 | 393%↑ |
| 6 | Median Number of Comments | 0%* | — | 1→2 | 3%↑ |
| 7 | Familiarity between the Invited Reviewer and the Patch Author | 0%* | — | 0→6 | 2%↑ |
| 8 | Number of Concurrent Reviews | 0%° | — | 12→103 | -1%↓ |
| 9 | Core Member Status | 0%° | — | 0→1 | -4%↓ |
| 10 | Number of Received Review Invitations | 0%° | — | 19→382 | 0% |
| 11 | Patch Size | 0%° | — | 13→489 | 0% |
| **Qt** | | | | | |
| 1 | Review Participation Rate | 60%* | 7%* | 74%→92% | 173%↑ |
| 2 | Reviewer Code Authoring Experience | 22%* | 1%* | 0→0.25 | 145%↑ |
| 3 | Number of Remaining Reviews | 5%* | — | 1→9 | -21%↓ |
| 4 | Familiarity between Invited Reviewer and Patch Author | 5%* | 5%* | 0→12 | -31%↓ |
| 5 | Number of Concurrent Reviews | 4%* | — | 8→39 | 21%↑ |
| 6 | Reviewer Reviewing Experience | 1%* | — | 0→0.006 | 1%↑ |
| 6 | Patch Author Reviewing Experience | 1%* | — | 0→0.004 | -1%↓ |
| 7 | Core Member Status | 1%* | — | 0→1 | 31%↑ |
| 8 | Number of Received Review Invitations | 1%* | — | 59→850 | -8%↓ |
| 9 | Patch Author Code Authoring Experience | 0%° | — | 0.07→0.6 | 4%↑ |
| 10 | Patch Size | 0%° | — | 4→70 | 0% |
| 10 | Median Number of Comments | 0%° | — | 0→88 | -63%↓ |

Statistical significant: * $p<0.001$ in more than 90% of the bootstrap samples; °otherwise
—: Nonlinear degrees of freedom are not allocated.

an indicator for the future participation decision of a reviewer without constructing a prediction model. However, our models also show that other factors (e.g., code authoring experience, familiarity between the reviewer and the patch author) also play a role. Hence, a prediction model may help practitioners to better select a reviewer while considering those factors.

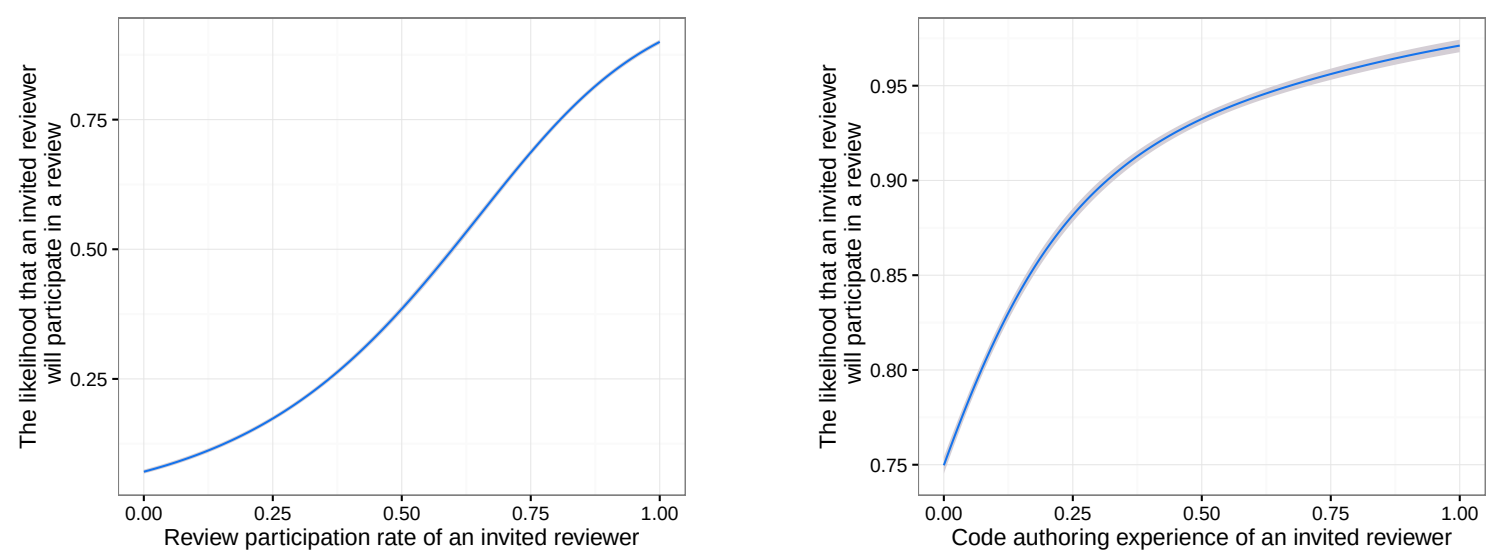

(a) Review participation rate in the Android model

(b) Code authoring experience of an invited reviewer in the Qt model

Fig. 9: The direction of the nonlinear relationships between the independent variable and the likelihood that an invited reviewer will participate in a review. The light grey area shows the 95% confidence interval.

**Observation 6 — Code authoring experience of an invited reviewer shares an increasing relationship with the likelihood that an invited reviewer will participate in a review.** Table 6 shows that the code authoring experience of an invited reviewer accounts for the second largest proportion of Wald $\chi^2$ in three of our four models, suggesting that the code authoring experience of an invited reviewer is second mostly associated with the participation decision in the three studied systems.

Figure 9(b) shows the direction of the relationship between the code authoring experience of an invited reviewer and the likelihood that an invited reviewer will participate in a review in the Qt model. Table 6 also shows that when the authoring experience of an invited reviewer increases from 0 to 0.19 in Android model, the likelihood increases by 53%. Similarly, when the authoring experience of an invited reviewer increases from 0 to 0.05, 0 to 0.08 and 0 to 0.25 in LibreOffice, OpenStack and Qt models respectively, the likelihood increases by 9%, 16% and 146%. These results indicate that an invited reviewer who has more authoring experience on the modules that are impacted by a patch is more likely to participate that patch.

To further understand the relationships corresponding to a core member status of a reviewer, we construct two more models for each studied dataset. One is the prediction model that includes only instances where the core member status is TRUE (i.e., a reviewer is a core member). The other prediction model that includes only instances where the core member status is FALSE (i.e., a reviewer is a non-core member). Figure 10 shows the direction of the relationships between the code authoring experience of an invited reviewer and the likelihood that an invited reviewer will participate in a review corresponding to the core member status. We observe that the differences in the likelihood to participate of core reviewers and non-core reviewers are small (i.e., 1%-10%). These results suggest that a core member status of invited reviewers has a weak relationship with the participation decision. Instead, inviting reviewers that have related experience with the modules that are impacted by the patch results in a higher likelihood that they will participate in the review.

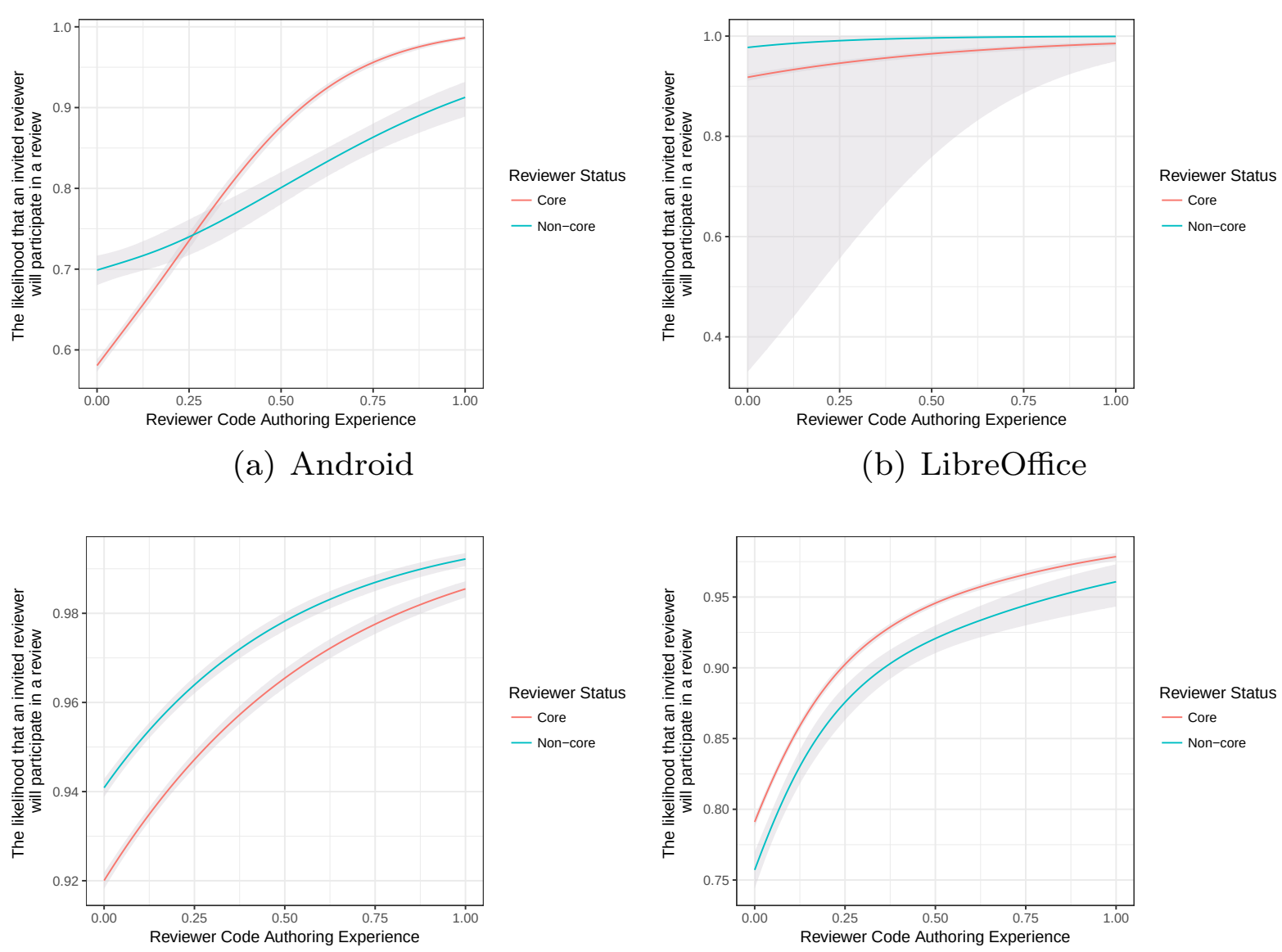

(a) Android (b) LibreOffice

(c) OpenStack (d) Qt

Fig. 10: The direction of the relationships between the code authoring experience of an invited reviewer and the likelihood that an invited reviewer will participate in a review corresponding to core member status. The light grey area shows the 95% confidence interval.

Table 6 shows that a core member status of an invited reviewer is ranked ninth for the Android & OpenStack models, eleventh for the LibreOffice model and seventh for the Qt model. In particular, the Wald $\chi^2$ of the core member status of an invited reviewer is statistically significant in the Android and Qt models. These results suggest that the core member status share a weak relationship to the participation decision of a reviewer. These results also highlight that despite the privilege of providing approval review of core reviewers, every reviewer shares a similar responsibility in reviewing a patch.

*The review participation rate is the most influential factor on the participation decision of an invited reviewer. In addition, the code authoring experience of an invited reviewer is also an influential factor on the participation decision of an invited reviewer (Observations 5-6).*

## 6 Practitioner Survey

To better gain insights into the review participation decision, we additionally conduct an online survey with the Android, OpenStack, and Qt developers. The survey questions consist of four parts:[14]

1. The respondent's demographic background,
2. Reviewer selection practices (i.e., how do patch authors select a reviewer for a patch),
3. Review participation decisions (i.e., what is the most likely reason that reviewers did not respond to the review invitation), and
4. Opinion on our study results (i.e., whether the respondents agree with our six empirical observations)

For our survey, we select the Android, Qt, and OpenStack developers with the following criteria: (1) developers who have been committed or commented patches in the last 365 days, (2) developers who were invited for a review more than 50 patches, and (3) developers who did not respond to more than 20% of the review invitations. We then sent out our survey questions to 130 Android developers, 98 OpenStack developers, and 110 Qt developers via emails. We describe our approach of retrieving the developer email addresses in a replication package.[15]

The survey was open for three weeks (from November 6 to November 26, 2017). We received 26 responses (8% of the 333 emails). We now present our survey results, which are grouped into four parts.

### 6.1 Survey Response Overview

24 of 26 respondents (92%) are both a reviewer and a patch author. There are 18 respondents who have contributed to the studied system for more than four years, while the other 6 respondents have contributed for more than two years. 25 respondents are a patch author where 40% of them submitted less than 5 patches per month, 32% of the patch authors submitted 5 to 15 patches per month, and 28% of the patch authors submitted 16 to 30 patches per month. 25 respondents are a reviewer where 64% of them responded to more than 50% of the review invitations, while 24% of them responded to 11% to 25% of the review invitations. In addition, 15 of the 25 reviewer respondents (60%) are a core reviewer. Since our studied datasets contain 7,496 unique developers across three software systems, our survey results have a margin of error of ±19.19% at the 95% confidence level.

### 6.2 Reviewer Selection Practices

Figure 11 shows the survey responses of how a patch author select a reviewer for a patch. 18 respondents who are a patch author report that they select a reviewer who committed or reviewed prior patches that impact the same module as their patches. 15 of the 25 patch authors also select a reviewer who often reviewed their patches in the past. In addition, 6 of the 25 patch authors report that they select a reviewer who is known to be very responsive. This result is consistent with our rationale for the participation rate metric, i.e., a high rate of review participation may indicate that the reviewer is active in the system (see Table 2). Moreover, our results in Table 6 also show that the participation rate is mostly associated with the likelihood that an invited reviewer will participate in a review. Nevertheless, the survey responses show that the respondents tend to consider several factors in addition to the participation rate metric. Hence, we believe that using a prediction model with a holistic view of several factors would help patch authors to better select a reviewer rather than using the past experience of the patch authors.

[14] We provide a full list of questions online at `https://goo.gl/forms/Du48JXAsbBhKSeSx2`.
[15] `https://github.com/sruangwan/replication-human-factors-code-review/`

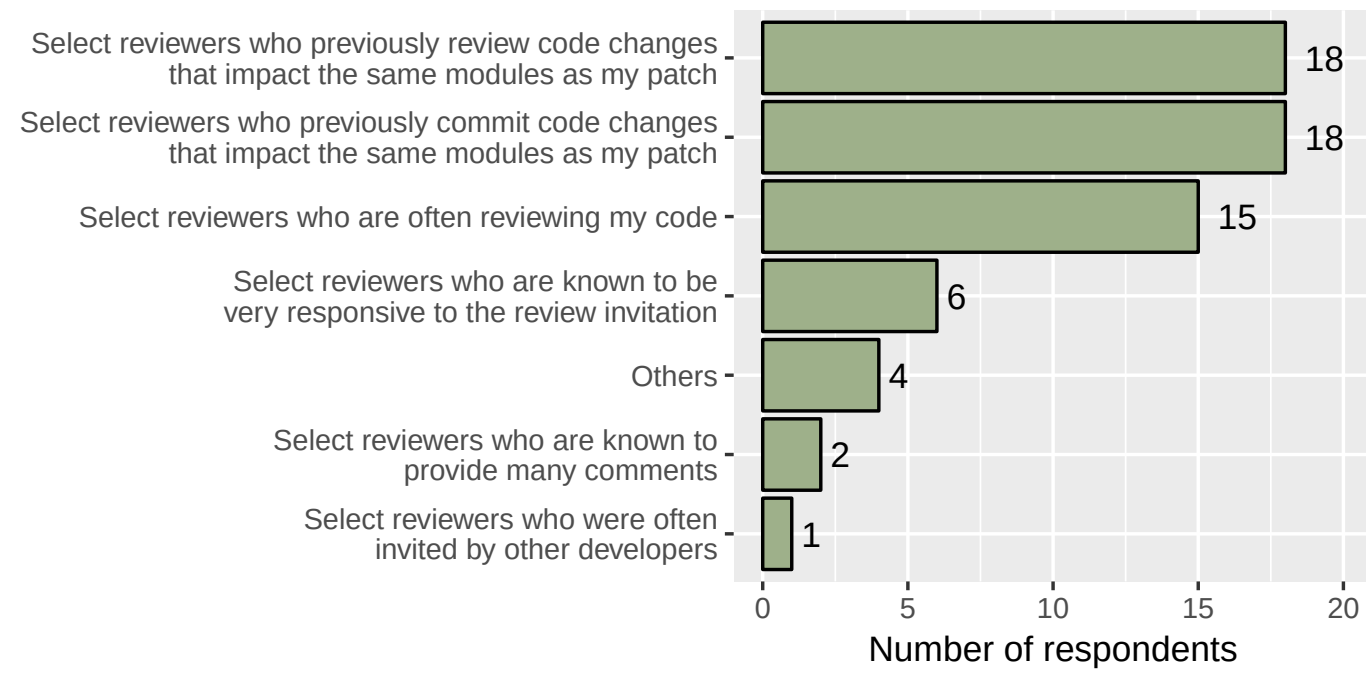

Fig. 11: Survey Responses: How do patch authors select a reviewer for a patch.

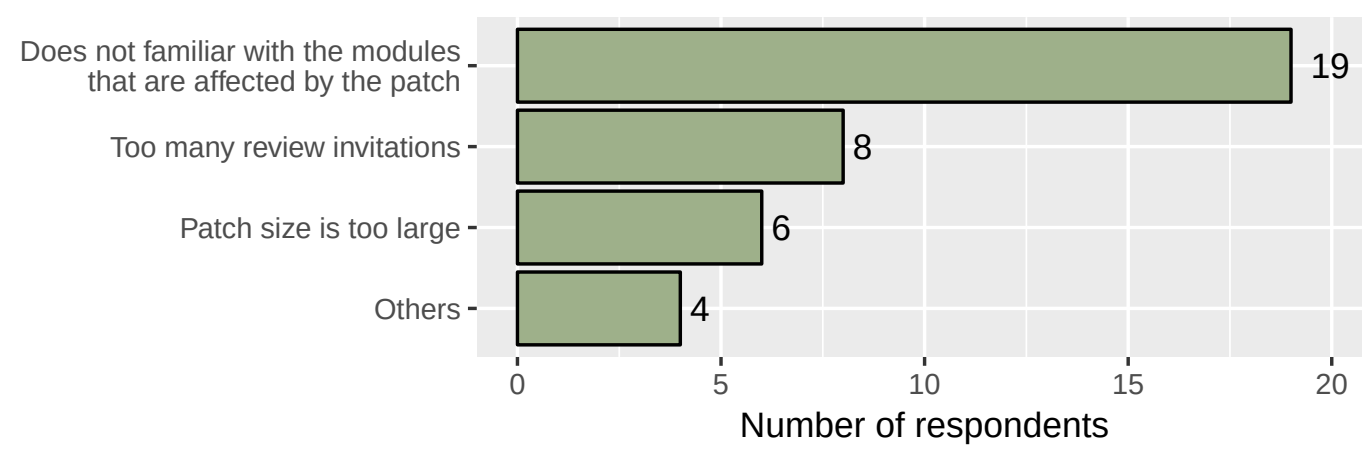

Fig. 12: Survey Responses: What is the most likely reason to not respond to the review invitation.

### 6.3 Review Participation Decisions

Figure 12 shows the survey responses of what is the most likely reason that reviewers did not respond to the review invitation. 19 of the 25 respondents who are reviewers reported that they did not respond to a review invitation because they were not familiar with the modules impacted by the patch. This result complements the intuition of the prior work that a reviewer who is familiar with the code in a patch is more likely to give a better review than others (Balachandran, 2013; Thongtanunam et al, 2015b; Xia et al, 2015; Yu et al, 2014; Zanjani et al, 2016).

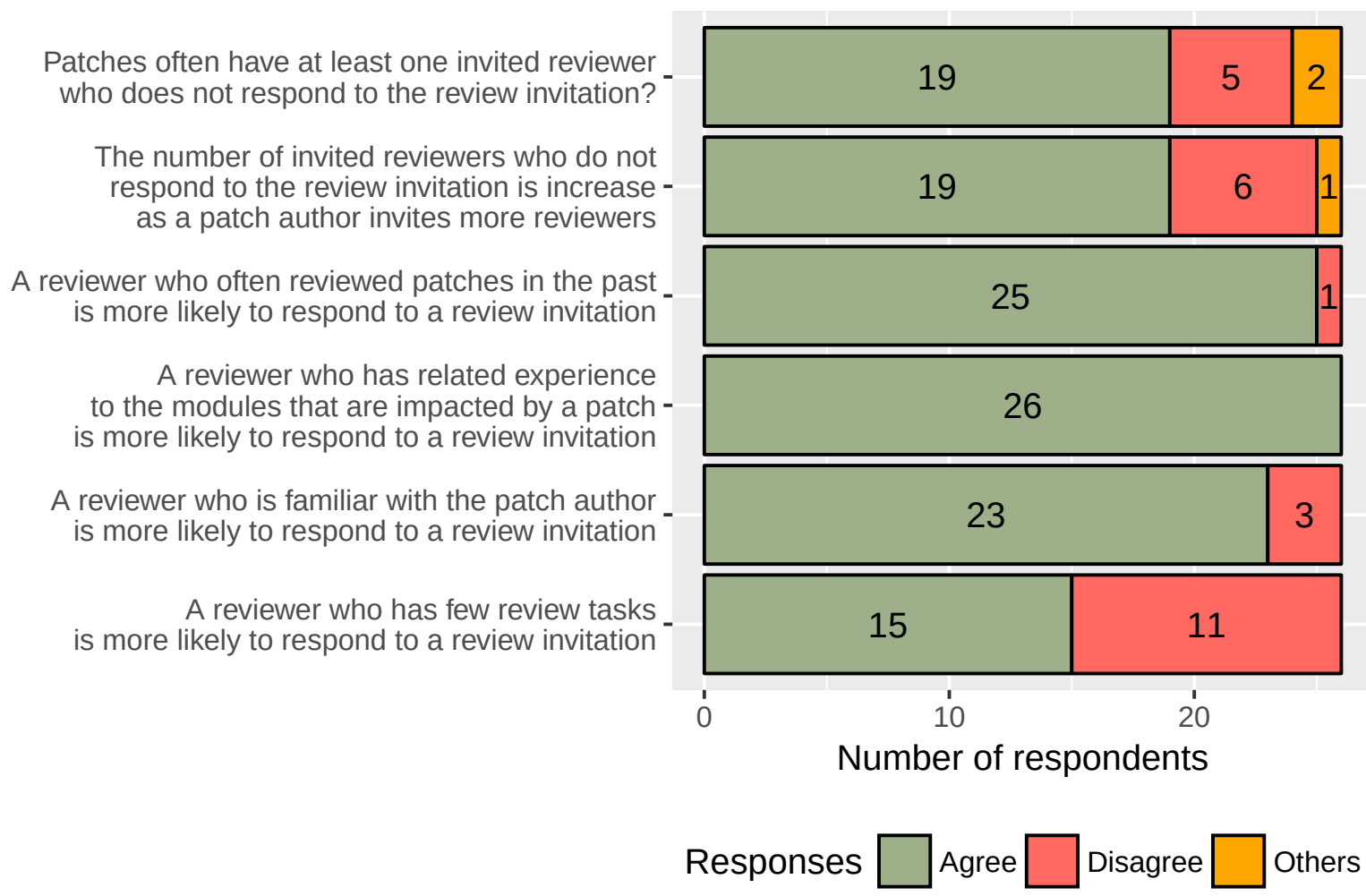

Fig. 13: Survey responses: Whether respondents agree with the empirical observations of our work.

This result is also consistent with our Observation 6 that the reviewer code authoring experience is highly associated with the participation decision. Moreover, 8 of the 25 reviewers also report that they did not respond to the review invitation because they received too many review invitations. This result is consistent with our results in Table 6 where the number of remaining reviews shares a significant relationship with the participation decision.

## 6.4 Opinion on Study Results

Figure 13 shows the survey responses of whether the respondents agree with our empirical observations. 19 of the 26 respondents agree that patches often have at least one invited reviewer who does not respond to the review invitation (i.e., Observation 1). 19 of the 26 respondents also agree that the number of unresponded review invitations increases as a patch author invites more reviewers (i.e., Observation 2). These results suggest that the respondents agree with our findings that patches often suffer from unresponded review invitations.

While Figure 11 shows that 6 of the 25 patch authors consider the responsiveness when selecting a reviewer, Figure 13 shows that most of the respondents (25 of the 26 respondents) agree with our Observation 5 that a reviewer who often reviewed patches in the past (i.e., the review participation rate) is more likely to respond to a review invitation. 15 of the 26 respondents also agree that a reviewer who has fewer review invitations in queue is more likely to respond to a review invitation. Furthermore, 23 of the 26 respondents agree that a reviewer who is familiar with the patch author is more likely to respond to a review invitation. These results suggest that the respondents agree that our uncovered human fac-tors and social interaction are important in determining whether or not a reviewer will participate a review.

All of the respondents also agree with our Observation 6 that a reviewer who has related experience to the modules impacted by a patch is more likely to respond to a review invitation. This result suggests that the respondents agree that experience of reviewers is one of the important factors of the participation decision.

In addition, we asked an open-ended question for opinion on unresponded review invitations. Several patch authors acknowledged that reviewers did not respond because they are busy. This finding is consistent with a survey study of Lee et al (2017) who find that many of the one-time patch authors acknowledged that the unresponsiveness of reviewers is in part due to the amount of workload. One of the respondents also raised a concern on the unresponded review invitations: *"... Anyhow, I feel people should take reviewer responsibilities quite seriously, even if reviewing other people patches is not that much fun."*. Another respondent also shared an opinion that reviewer participation should be investigated: *"... I think there's an interesting social dynamic in how some people shy away from +2's in such a setup, and am wondering how to change that."*. These responses support our motivation that a better understanding of the factors associated with participation decision would help software development teams to develop better strategies for the code review process.

## 7 Discussion

In this section, we further discuss our findings and provide a practical suggestion.

### 7.1 The Participation Decision of Reviewers

Observation 1 shows that 16%-66% of patches have at least one reviewer who did not respond to the review invitation, indicating that patches often have reviewers who did not respond to the review invitation. Observation 2 shows that the number of invited reviewers shares an increasing relationship with the number of reviewers who did not respond to the review invitation. These results suggest that the more reviewers were invited to a patch, the more likely that the invited reviewers will not respond to the review invitation. One possible reason is the tragedy of commons, where an invited reviewer did not review a patch since there were many invited reviewers and the patch still has a chance to get reviewed by other invited reviewers (Hardin, 1968). Another possible scenario is broadcasting review invitations (Rigby and Storey, 2011). The invited reviewers did not respond to the review invitation since other reviewers who have similar expertise already reviewed the patch. Furthermore, Kononenko et al (2015) find that the number of invited reviewers is associated with the number of defects. Therefore, a patch author should not invite many reviewers.

### 7.2 Human Factors

Observations 3 and 4 show that human factors increase the performance of our models that predict whether or not an invited reviewer will participate in a review.

These results suggest that in addition to experience and technical factors, patch authors should understand that human-related factors can have an impact on the participation decision of reviewers. As the approach of inviting more reviewers to increase review participation is becoming less efficient when the number of invited reviewers is increasing, a better understanding of factors playing a role in code review process can be helpful.

Observation 5 shows that the review participation rate of an invited reviewer accounts for the most influential factor on the participation decision of reviewers. More specifically, the review participation rate shares an increasing nonlinear relationship with the likelihood that an invited reviewer will participate in a review. This finding suggests that an active reviewer who has been actively responded to a review in the past is more likely to respond to the review invitation.

In addition to the review participation rate, other human factors also share a strong relationship with the likelihood that an invited reviewer will participate in a review. For example, Table 6 shows that the familiarity between the invited reviewer and the patch author is ranked tenth, ninth, seventh, and fourth by explanatory power in the Android, LibreOffice, OpenStack, and Qt models respectively. Furthermore, the familiarity between the invited reviewer and the patch author is also statistically significant in three of four studied systems. This result arrives at the similar finding of Kononenko et al (2016), who find that the relationship or trust between a reviewer and patch author can have an impact on the review outcome.

Moreover, the reviewer workload (i.e., the number of concurrent reviews and the number of remaining reviews) shares a statistically significant relationship with the likelihood. In particular, the number of concurrent reviews is ranked seventh, fifth, eighth, and fifth by explanatory power in the Android, LibreOffice, OpenStack, and Qt models respectively, while the number of remaining reviews is ranked fifth, third and third in the Android, LibreOffice, and Qt models respectively. This result also complements to the findings of Baysal et al (2013), who find that the review queue of reviewers has an impact on the review timeliness and the review outcome.

### 7.3 Technical Factors

Observation 6 shows that the code authoring experience of an invited reviewer accounts for the second most influential factor on the participation decision of reviewers. More specifically, the code authoring experience of an invited reviewer shares an increasing relationship with the likelihood that an invited reviewer will participate in a review. Although the code authoring experience does not contribute as large explanatory power ($\chi^2$) as the review participation rate, the code authoring experience does contribute a relatively large contribution in the Android and Qt models. The result is consistent with the intuition of prior work (Balachandran, 2013; Thongtanunam et al, 2015b; Xia et al, 2015; Yu et al, 2014), i.e., a reviewer is more likely to participate in a review of the patch if the reviewer has related experience with the patch.

Thongtanunam et al (2016a) find that patch size shares a relationship with the likelihood that a patch will suffer from poor review participation. However, Table 6 shows that patch size has an insignificant impact on the likelihood that an invited reviewer will participate in a review. This finding suggests that although patch size impacts a patch whether or not it will suffer from poor review participation, patch size has a very small impact on the participation decision of an individual reviewer. We think the reason is there are other factors that reviewers consider when they decide to participate in a review. For example, a reviewer may not participate in a review even though the patch size is small (i.e., easy to review) because the reviewer has no related experience with the modules that are impacted by the patch.

Furthermore, we find that the code authoring experience of patch author and the reviewing experience of patch author share a relationship with the likelihood that an invited reviewer will participate in a review. In other words, the experience can also indicate the reputation of a patch author, i.e., the more patches the author made to the system, the more well-known the author is. Bosu and Carver (2014) have shown that patch authors with high reputation (i.e., core developers) tend to receive quicker first feedback on their patches than patch authors with the lower reputation (i.e., peripheral developers). These findings suggest that reviewers are more likely to participate in a patch that is made by the patch author with high experience.

### 7.4 Practical Suggestions and Recommendations for Future Work

We construct a prediction model that leverages human factors, experience, and technical factors to predict whether or not an invited reviewer will participate in a review. Our results show that human factors should be considered in addition to technical and experience factors when inviting reviewers.

Practitioners may simply use a single metric (e.g., the reviewer experience, the review participation rate) as an indicator for the future participation. However, our results show that considering a single metric may not be sufficient in a prediction since other metrics also share a significant relationship to the participation decision of an invited reviewer. For example, one might count the number of commits to identify reviewers, however those reviewers may not respond to the review invitation due to a high workload. Similarly, a reviewer who has a high review participation rate but has little reviewing or authoring experience on the modified modules is less likely to respond the review invitation. Moreover, solely considering a single metric when inviting reviewers may lead them to be overwhelmed by review invitations. Therefore, a prediction model that has a holistic view of both technical and human factors would help patch authors to better select a reviewer than simply using a single metric.

Furthermore, our results show that human factors (e.g., review workload and familiarity between reviewers and patch authors) share a significant relationship with the participation decision of an invited reviewer. This finding could complement a reviewer recommendation approach of the prior work (Balachandran, 2013; Thongtanunam et al, 2015b; Xia et al, 2015; Yu et al, 2014; Zanjani et al, 2016). In other words, future work of reviewer recommendation should consider human factors in order to better find a reviewer.

To demonstrate how our prediction models can help practitioners, we use our models that include human factors to estimate the likelihood that the invited reviewers will respond to the review invitation. We then measure the top-$k$ accuracy ($k = 1, 2, 3$), i.e., a proportion of patches where an invited reviewer, who has the highest participation likelihood estimated by our models, will respond to the review invitation. Table 7 shows the top-$k$ accuracy of our models where the top-$k$ accuracy is ranging from 0.91 to 1. These results indicate that our models can accurately recommend reviewers who will participate in a review. One possible usage scenario is that a patch author (or a reviewer recommendation tool) first lists the potential reviewers, then invites only the reviewers who are more likely to respond to the review invitation based on the estimation of our models.

Another benefit of using our models is to reduce the number of review invitations. Reducing the number of review invitations may help practitioners increase review quality since Kononenko et al (2015) report that the number of invited reviewers is associated with the review bugginess. To demonstrate this benefit, we measure a proportion of unresponded review invitations that are predicted by our models. We find that 3% (LibreOffice and OpenStack) to 31% (Android) of the review invitations are predicted as unresponded review invitations by our models, implying that these 3% to 31% of review invitations are not necessary to be made since reviewers are less likely to respond the review invitations. The negative predictive value (i.e., $\frac{\text{\#Correctly predicted as unresponded review invitations}}{\text{\#Predicted as unresponded review invitations}}$) of our models is also relatively high (66% to 75%), indicating that our models accurately identify the unresponded review invitations. Based on this analysis, the number of review invitations can be reduced by 3% to 31% if a patch author did not invite reviewers as suggested by our models.

### 7.5 Differences between OSS Communities

Our observations 1 and 2 show that patches in the LibreOffice and OpenStack datasets tend to less suffer from the non-responding reviewers than the other two studied datasets. In particular, the LibreOffice and OpenStack datasets have less percentage of patches that have at least one invited reviewer who did not respond to the review invitation than the other two datasets. Additionally, the correlation between the number of invited reviewers and the number of reviewers who did not respond to the review invitation is the lowest in the OpenStack dataset, while it is the second lowest in the LibreOffice dataset. One possible explanation for the different results between the systems is the activeness of reviewers in the systems. Figure 14 shows the distribution of participation rate of reviewers. At the median, the LibreOffice reviewers typically respond to 90% of patches that they were invited. Similarly, the OpenStack reviewers typically respond to 92% of patches that they were invited. However, the Android and Qt reviewers typically respond to 60% and 78% of patches, respectively. This result indicates that the LibreOffice and OpenStack systems tend to have more active reviewers than the other systems.

Table 7: Top-$k$ accuracy ($k = 1, 2, 3$) of our prediction models

| Top-k Accuracy | Android | LibreOffice | OpenStack | Qt |
|---|---|---|---|---|
| Top-1 | 0.91 | 0.98 | 0.99 | 0.95 |
| Top-2 | 0.95 | 0.99 | 1.00 | 0.99 |
| Top-3 | 0.95 | 0.99 | 1.00 | 1.00 |

In addition to the activeness, prior work also finds that developers in OpenStack have the highest closeness centrality while developers in Android have the lowest closeness centrality (Yang et al, 2016b). The closeness centrality can be positively associated to the closeness of the people in the community (Freeman, 1978). Therefore, having a strong community can potentially be the reason that makes OpenStack system less suffer from the non-responding reviewers.

The diversity of participating organizations in the software systems may also play a role in the participation decision of reviewers. To investigate this, we determine an organization of developers in the studied systems using the domain name in developer email addresses. We then count the number of developers of each organization. We find that there are 202, 123, 689, and 349 organizations participating in Android, LibreOffice, OpenStack, and Qt, respectively. Figure 15 shows a proportion of organizations in each studied system. We observe that developers from Google is the majority (25%) in the Android system. Similarly, the number of developers in the leading teams of Qt, which are from Nokia, Digia, and The Qt Company, accounts for 22%.[16,17] On the other hand, although the OpenStack project was led by Rackspace, the number of developers from Rackspace accounts for only 5%.[18] Moreover, OpenStack is known to be supported by more than 500 companies as of 2018.[19] The number of developers in the leading team of LibreOffice accounts for only 3%. Furthermore, LibreOffice defines its software as a community-driven and developed software.[20] These results suggest that the developers in LibreOffice and OpenStack are more diverse than Android and Qt, implying that LibreOffice and OpenStack datasets tend to less suffer from the non-responding reviewers may be in part due to the diversity of developers.

## 8 Threats to Validity

We now discuss the threats to validity of our study.

[16] `https://www.theguardian.com/technology/2012/aug/09/nokia-sells-qt-software-business/`
[17] `https://www.zdnet.com/article/qt-hot-potato-spun-out-from-digia-into-fourth-home/`
[18] `https://www.openstack.org/blog/?p=1`
[19] `https://www.openstack.org/foundation/companies/`
[20] `https://www.libreoffice.org/about-us/who-are-we/`

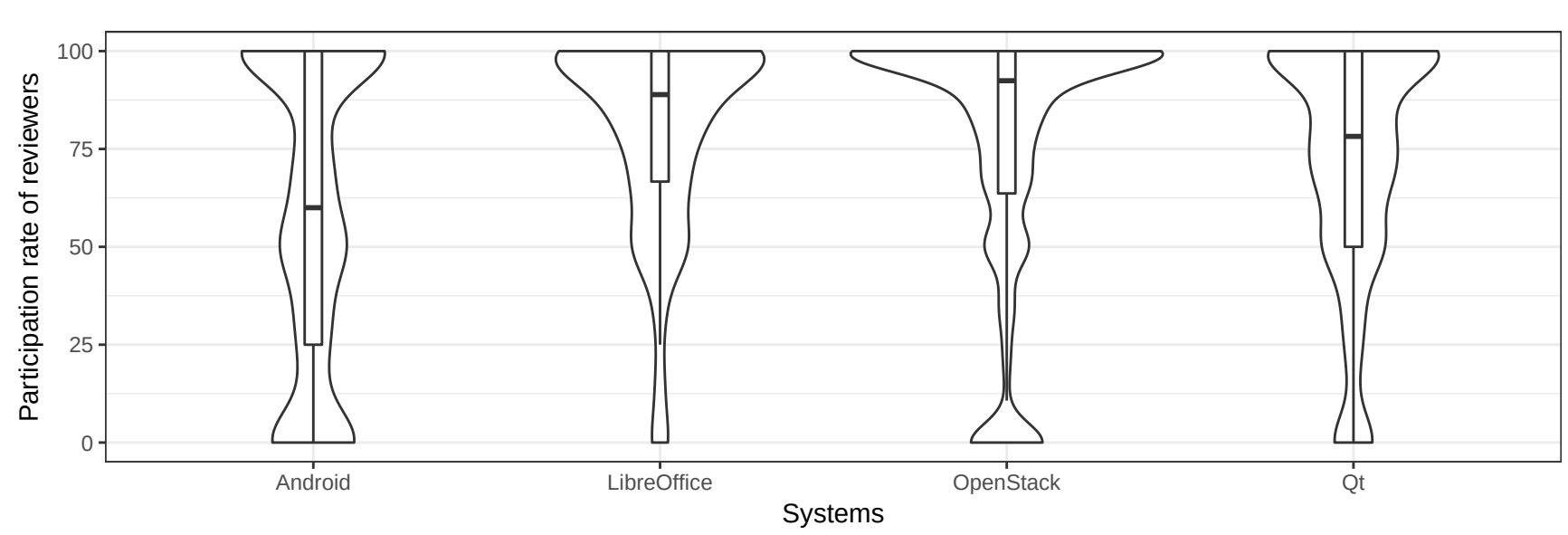

Fig. 14: The distributions of participation rate of reviewers.

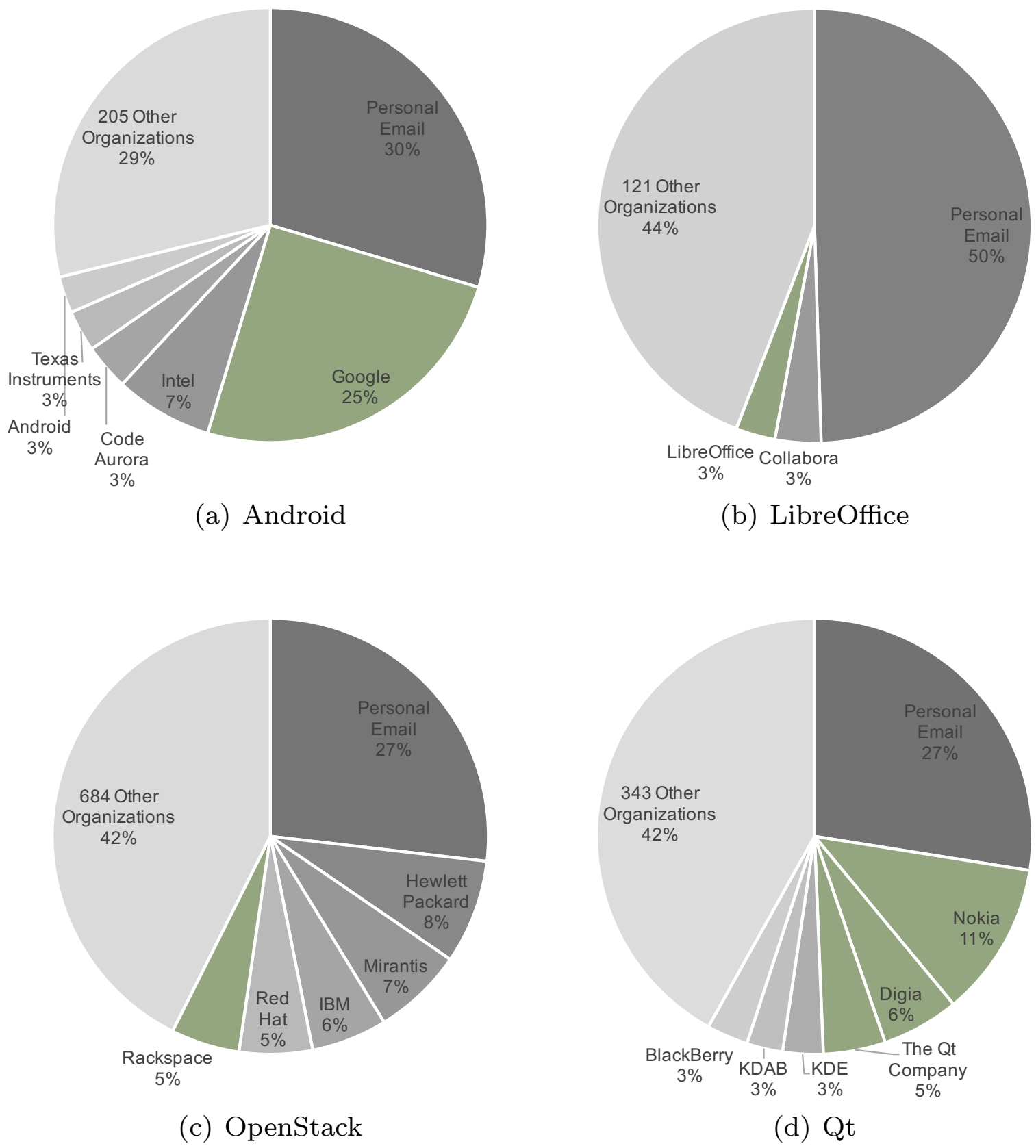

Fig. 15: A pie chart of the proportion of organizations participating in each studied system. The green slices are the proportion of developers in the leading teams.

## 8.1 Construct Validity

We compute our studied metrics at the creation time of patches. Unfortunately, the Gerrit code review tool does not record when the author invites a reviewer. Hence, we must rely on this heuristic and assume that all reviewers are invited at the same time as the creation time of patches. Be able to analyze the exact time reviewers were invited will allow us to analyze the code review practices while aware of the time component. For example, a patch author invites two reviewers at the creation time of a patch, but only one of them responds to the invitation. To increase the review participation, the patch author then invites two more reviewers.

We measure workload of an invited reviewer based on a heuristic that the invited reviewer will review the patch from the creation time until the patch reaches a final decision. However, there are likely cases where reviewers only review a patch for a part of this time frame. Unfortunately, the Gerrit code review tool does not record the time that reviewers actually spent reviewing a patch. Therefore, we must also rely on this heuristic (see the calculation of review workload metrics in Section 4). Be able to analyze the time frame that each reviewer truly spend reviewing will enable more accurate values of review workload metrics.

We assume that once a reviewer became a core member (i.e., a reviewer had provided a review score of +2 or -2 in the past), the core member status will not be reverted back to a non-core member status. However, van Wesel et al (2017) find that the core member status may be reverted back to a non-core member status based on the reviewing activities in the past. To address this possible threat, we check the voting range that was actually permitted for reviewers in each patch of the Android dataset using Gerrit REST API.[21] In other words, if a reviewer that had a permission to vote a review score of +2 or -2 for a patch, that reviewer should have a core member status during the review of that patch. Based on this ground-truth data, our heuristic (i.e., observing the provided review scores) can correctly identify the core member status for 75% of instances in the Android dataset. Unfortunately, such voting permission information is not publicly available in the LibreOffice, OpenStack, and Qt systems. Based on the result of the Android dataset, we believe that we can rely on our heuristic to identify a core member status. Nevertheless, a more accurate approach of identifying a core member status may further strengthen our findings.

The prediction models of LibreOffice, OpenStack and Qt achieve a relative low recall compare to the Android model. One technique that could improve the recall of our models is rebalancing the data. However, in addition to model performance, another goal of this study is to examine the signals that can relate to the review participation decision of a reviewer. Tantithamthavorn et al (2017a) point out that rebalancing techniques have a negative impact on the interpretation of regression models. Therefore, in this study, we build our models to fit the original data rather than rebalancing data in order to truly understand the relationship between the studied factors and the participation decision of a reviewer. Nevertheless, we examine the recall of our models after rebalancing the data using random over-sampling examples (ROSE) technique (Menardi and Torelli, 2014). We find that the recall value increases by 0.55, 0.34 and 0.40 points for the LibreOffice, OpenStack and Qt models, respectively.

### 8.2 External Validity

We perform a study on four open source software systems that use the Gerrit code review tool, which may limit the generalizability of our results. Additionally, we find that there is a possibility that the same metric performs differently for different systems (e.g., the reviewing experience of an invited reviewer). However, the goal of this study is not to define a wide range theory that holds true for every project. Instead, our key contribution of this study is to show that in some settings of code review process, the human factors can play an important role. Nonetheless, we facilitate future work with a replication package of R scripts.[22] Future work should expand the study to include other software systems and code review tools to establish the validity of our findings in other contexts.

[21] https://gerrit-review.googlesource.com/Documentation/rest-api-changes.html#approval-info

[22] https://github.com/sruangwan/replication-human-factors-code-review/

### 8.3 Internal Validity

We identify whether or not an invited reviewer participated in a review of a patch using a review score and comments that are posted in the patch. However, it is possible that the invited reviewers perform code review through other communication media such as in-person discussion (Bacchelli and Bird, 2013; Guzzi et al, 2013), a group IRC (Shihab et al, 2009) or a mailing list (Rigby et al, 2008). Since we identify the participation decision based on comments and review score, performing code review outside of the platform may lead to an inaccurate participation decision. Nonetheless, we perform our study on the systems that actively perform a code review on the Gerrit code review tool, which should capture the majority of the code review activity.

We assume that reviewers are invited at the beginning of the code review process. However, it is possible that reviewers were invited at different points in time. For example, a patch author invites one reviewer but that reviewer did not respond. Then, the patch author invites other reviewers. Unfortunately, the Gerrit system does not record time when each reviewer was invited. Therefore, we have to rely on this assumption. A real-time data collection may help future work to better understand the reviewer invitation process.

There is a chance that software development policies confound our findings. For example, the relationship of code authoring experience (i.e., code ownership) and the participation decision may be affected by a policy that patches have to be approved by a core reviewer who usually has high experience. We control this concern by including a reviewer status (i.e., core or non-core reviewer) to our studied metrics. Our results show that the reviewer status does not have much impact on the participation decision, suggesting that this requirement policy does not impact the participation decision. Another example is that LibreOffice and OpenStack systems tend to have more active reviewers than the other systems. However, we cannot find any special policy of LibreOffice and OpenStack systems that potentially causes this outcome.

## 9 Conclusions

The flexibility of Modern Code Review (MCR) process allows reviewers to decide whether or not to participate in a review. Such a practice becomes one of the main challenges of MCR process. Despite the impact of poor review participation that several studies have found (Bavota and Russo, 2015; Bettenburg et al, 2015; McIntosh et al, 2014), little is known about the current practices of reviewer participation. Moreover, the factors (especially the human factors) that can influence the participation decision of reviewers remain largely unexplored. In this paper, we analyze descriptive statistics of the number of reviewers who did not respond to the review invitation of patches. We then construct prediction models to determine the likelihood of the participation decision of reviewers, and to understand the factors that influence the participation decision. Through a case study of the Android, LibreOffice, OpenStack, and Qt systems, we empirical study 230,090 patches, we make the following observations:

- A large number of patches (i.e., 16%-66%) have at least one invited reviewer who did not respond to the review invitation. Moreover, the number of invited reviewers has a medium to large correlation with the number of reviewers who did not respond to the review invitation (Observations 1-2).
- Our prediction models that include human factors outperform the baseline models with an AUC value of 0.82-0.89, a Brier score of 0.06-0.13, a precision of 0.68-0.78, a recall of 0.24-0.73, and an F-measure of 0.35-0.75. These results suggest that human factors play an important role in determining the likelihood of the participation decision of reviewers (Observations 3-4).
- The review participation rate of an invited reviewers shares a strong increasing relationship with the likelihood that an invited reviewer will participate in a review. Additionally, the code authoring experience of an invited reviewer also shares an increasing relationship with the likelihood (Observations 5-6).

We believe that our results and observations shed the light of understanding the current practices of reviewer participation which may lead to poor review participation. Our results also highlight the importance of human factors which have an impact on the participation decision of reviewers. Patch authors should take human factors into the consideration when inviting reviewers for a new patch because it may increase the likelihood that an invited reviewer will participate in a review. To facilitate future work, we provide a replication package of R scripts online.[23]

**Acknowledgments**

This research was partially supported by JSPS KAKENHI Grant Number 16J02861 and 17H00731, and Support Center for Advanced Telecommunications (SCAT) Technology Research, Foundation. We would also like to thank Dr. Chakkrit Tantithamthavorn for his insightful comments and the survey participants for their time.

**References**

Ackerman AF, Buchwald LS, Lewski FH (1989) Software inspections: an effective verification process. IEEE Software 6(3):31–36

Armstrong F, Khomh F, Adams B (2017) Broadcast vs. Unicast Review Technology: Does It Matter? In: Proceedings of the 10th International Conference on Software Testing, Verification and Validation (ICST), pp 219–229

Bacchelli A, Bird C (2013) Expectations, Outcomes, and Challenges of Modern Code Review. In: Proceedings of the 35th International Conference on Software Engineering (ICSE), pp 712–721

Balachandran V (2013) Reducing Human Effort and Improving Quality in Peer Code Reviews using Automatic Static Analysis and Reviewer Recommendation. In: Proceedings of the 35th International Conference on Software Engineering (ICSE), pp 931–940

[23] `https://github.com/sruangwan/replication-human-factors-code-review/`

Bavota G, Russo B (2015) Four Eyes Are Better Than Two: On the Impact of Code Reviews on Software Quality. In: Proceedings of the 31st International Conference on Software Maintenance and Evolution (ICSME), pp 81–90

Baysal O, Kononenko O, Holmes R, Godfrey MW (2013) The Influence of Non-technical Factors on Code Review. In: Proceedings of the 20th Working Conference on Reverse Engineering (WCRE), pp 122–131

Beller M, Bacchelli A, Zaidman A, Juergens E (2014) Modern Code Reviews in Open-source Projects: Which Problems Do They Fix? In: Proceedings of the 11th Working Conference on Mining Software Repositories (MSR), pp 202–211

Bettenburg N, Hassan AE, Adams B, German DM (2015) Management of community contributions - A case study on the Android and Linux software ecosystems. Empirical Software Engineering (EMSE) 20(1):252–289

Bird C, Nagappan N, Murphy B, Gall H, Devanbu P (2011) Don't Touch My Code!: Examining the Effects of Ownership on Software Quality. In: Proceedings of the 19th ACM SIGSOFT Symposium and the 13th European Conference on Foundations of Software Engineering (ESEC/FSE), pp 4–14

Bosu A, Carver JC (2014) Impact of Developer Reputation on Code Review Outcomes in OSS Projects: An Empirical Investigation. In: Proceedings of the 8th International Symposium on Empirical Software Engineering and Measurement (ESEM), pp 33:1–33:10

Brier GW (1950) Verification of forecasts expressed in terms of probability. Monthly Weather Review 78(1):1–3

Carr DB, Littlefield RJ, Nichloson WL, Littlefield JS (1987) Scatterplot Matrix Techniques for Large N. Journal of the American Statistical Association (JASA) 82(398):424–436

Cliff N (1993) Dominance Statistics: Ordinal Analyses to Answer Ordinal Questions. Multivariate Behavioral Research 114(3):494–509

Cliff N (1996) Answering Ordinal Questions with Ordinal Data Using Ordinal Statistics. Multivariate Behavioral Research 31(3):331–350

Cohen J (1992) Statistical Power Analysis. Current Directions in Psychological Science 1(3):98–101

Croux C, Dehon C (2010) Influence functions of the Spearman and Kendall correlation measures. Statistical Methods & Applications (SMA) 19(4):497–515

Edmundson A, Holtkamp B, Rivera E, Finifter M, Mettler A, Wagner D (2013) An Empirical Study on the Effectiveness of Security Code Review. In: Proceedings of the 5th International Conference on Engineering Secure Software and Systems (ESSoS), pp 197–212

Efron B (1983) Estimating the Error Rate of a Prediction Rule: Improvement on Cross-Validation. Journal of the American Statistical Association (JASA) 78(382):316–331

Elish KO, Elish MO (2008) Predicting Defect-prone Software Modules Using Support Vector Machines. Journal of Systems and Software 81(5):649–660

Fagan ME (1976) Design and Code Inspections to Reduce Errors in Program Development. IBM Systems Journal 15(3):182–211

Fagan ME (1986) Advances in Software Inspections. Transactions on Software Engineering (TSE) 12(7):744–751

Fawcett T (2006) An introduction to ROC analysis. Pattern Recognition Letters 27(8):861–874

Foo KC, Jiang ZMJ, Adams B, Hassan AE, Zou Y, Flora P (2015) An Industrial Case Study on the Automated Detection of Performance Regressions in Heterogeneous Environments. In: Proceedings of the 37th International Conference on Software Engineering (ICSE), pp 159–168

Freeman LC (1978) Centrality in Social Networks Conceptual Clarification. Social Networks 1(3):215–239

Goeminne M, Mens T (2011) Evidence for the Pareto principle in Open Source Software Activity. In: Proceedings of the 1st International workshop on Model Driven Software Maintenance (MDSM) and 5th International Workshop on Software Quality and Maintainability (SQM), pp 74–82

Guzzi A, Bacchelli A, Lanza M, Pinzger M, van Deursen A (2013) Communication in Open Source Software Development Mailing Lists. In: Proceedings of the 10th Working Conference on Mining Software Repositories (MSR), pp 277–286

Hahn J, Moon JY, Zhang C (2008) Emergence of New Project Teams from Open Source Software Developer Networks: Impact of Prior Collaboration Ties. Information Systems Research 19(3):369–391

Hamasaki K, Kula RG, Yoshida N, Cruz AEC, Fujiwara K, Iida H (2013) Who Does What During a Code Review? Datasets of OSS Peer Review Repositories. In: Proceedings of the 10th Working Conference on Mining Software Repositories (MSR), pp 49–52

Hanley Ja, McNeil BJ (1982) The Meaning and Use of the Area under a Receiver Operating Characteristic (ROC) Curve. Radiology 143(4):29–36

Hardin G (1968) The Tragedy of the Commons. Science 162(3859):1243–1248

Harrell Jr FE (2002) Regression Modeling Strategies, 1st edn. Springer

Harrell Jr FE (2015a) Hmisc: Harrell Miscellaneous. `http://CRAN.R-project.org/package=Hmisc`

Harrell Jr FE (2015b) Regression Modeling Strategies, 2nd edn. Springer

Harrell Jr FE (2015c) rms: Regression Modeling Strategies. `http://CRAN.R-project.org/package=rms`

Hinkle DE, Wiersma W, Jurs SG (1998) Applied Statistics for the Behavioral Sciences, 4th edn. Houghton Mifflin Boston

Huizinga D, Kolawa A (2007) Automated Defect Prevention: Best Practices in Software Management. John Wiley & Sons

Kononenko O, Baysal O, Guerrouj L, Cao Y, Godfrey MW (2015) Investigating Code Review Quality: Do People and Participation Matter? In: Proceedings of the 31st International Conference on Software Maintenance and Evolution (ICSME), pp 111–120

Kononenko O, Baysal O, Godfrey MW (2016) Code Review Quality: How Developers See It. In: Proceedings of the 38th International Conference on Software Engineering (ICSE), pp 1028–1038

Lanubile F, Ebert C, Prikladnicki R, Vizcano A (2010) Collaboration Tools for Global Software Engineering. Software 27(2):52–55

Lee A, Carver JC, Bosu A (2017) Understanding the Impressions, Motivations, and Barriers of One Time Code Contributors to FLOSS Projects: A Survey. In: Proceedings of the 39th International Conference on Software Engineering (ICSE), pp 187–197

Liang J, Mizuno O (2011) Analyzing Involvements of Reviewers through Mining a Code Review Repository. In: Proceedings of the 21st International Workshop on Software Measurement and the 6th International Conference on Software Process and Product Measurement (IWSM-Mensura), pp 126–132

Mason CH, Perreault Jr WD (1991) Collinearity, Power, and Interpretation of Multiple Regression Analysis. Journal of Marketing Research (JMR) 28(3):268–280

McGraw G (2004) Software Security. Security & Privacy 2(2):80–83

McIntosh S, Kamei Y, Adams B, Hassan AE (2014) The Impact of Code Review Coverage and Code Review Participation on Software Quality: A Case Study of the Qt, VTK, and ITK Projects. In: Proceedings of the 11th Working Conference on Mining Software Repositories (MSR), pp 192–201

McIntosh S, Kamei Y, Adams B, Hassan AE (2016) An Empirical Study of the Impact of Modern Code Review Practices on Software Quality. Empirical Software Engineering (EMSE) 21(5):2146–2189

Menardi G, Torelli N (2014) Training and assessing classification rules with imbalanced data. Data Mining and Knowledge Discovery 28(1):92–122

Meyer B (2008) Design and Code Reviews in the Age of the Internet. Communications of the ACM 51(9):66–71

Mishra R, Sureka A (2014) Mining Peer Code Review System for Computing Effort and Contribution Metrics for Patch Reviewers. In: Proceedings of the 4th Workshop on Mining Unstructured Data (MUD), pp 11–15

Mukadam M, Bird C, Rigby PC (2013) Gerrit Software Code Review Data from Android. In: Proceedings of the 10th Working Conference on Mining Software Repositories (MSR), pp

45–48

Newson R (2002) Parameters behind "non-parametric" statistics: Kendall's tau, Somers' D and median differences. Stata Journal 2(1):45–64(20)

Rigby PC, Storey MA (2011) Understanding Broadcast Based Peer Review on Open Source Software Projects. In: Proceedings of the 33rd International Conference on Software Engineering (ICSE), pp 541–550

Rigby PC, German DM, Storey MA (2008) Open Source Software Peer Review Practices: A Case Study of the Apache Server. In: Proceedings of the 30th International Conference on Software Engineering (ICSE), pp 541–550

Rigby PC, Cleary B, Painchaud F, Storey MA, German DM (2012) Open Source Peer Review Lessons and Recommendations for Closed Source. IEEE Software

Rigby PC, German DM, Cowen L, Storey MA (2014) Peer Review on Open-Source Software Projects: Parameters, Statistical Models, and Theory. Transactions on Software Engineering and Methodology (TOSEM) 23(4):35:1–35:33

Sarle W (1990) The VARCLUS Procedure, 4th edn. SAS Institute, Inc.

Shihab E, Jiang ZM, Hassan AE (2009) Studying the Use of Developer IRC Meetings in Open Source Projects. In: Proceedings of the 25th International Conference on Software Maintenance (ICSM), pp 147–156

Spearman C (1904) The Proof and Measurement of Association between Two Things. The American Journal of Psychology (AJP) 15(1):72–101

Steinmacher I, Conte T, Gerosa MA, Redmiles D (2015) Social Barriers Faced by Newcomers Placing Their First Contribution in Open Source Software Projects. In: Proceedings of the 18th ACM Conference on Computer Supported Cooperative Work & Social Computing (CSCW), pp 1379–1392

Tantithamthavorn C, Hassan AE (2018) An experience report on defect modelling in practice: Pitfalls and challenges. In: Proceedings of the 40th International Conference on Software Engineering: Software Engineering in Practice (ICSE-SEIP), pp 286–295

Tantithamthavorn C, McIntosh S, Hassan AE, Ihara A, Matsumoto K (2015) The Impact of Mislabelling on the Performance and Interpretation of Defect Prediction Models. In: Proceedings of the 37th International Conference on Software Engineering (ICSE), pp 812–823

Tantithamthavorn C, McIntosh S, Hassan AE, Matsumoto K (2016) Comments on "Researcher Bias: The Use of Machine Learning in Software Defect Prediction". Transactions on Software Engineering (TSE) 42(11):1092–1094

Tantithamthavorn C, Hassan AE, Matsumoto K (2017a) The Impact of Class Rebalancing Techniques on the Performance and Interpretation of Defect Prediction Models. Under Review at Transactions on Software Engineering (TSE)

Tantithamthavorn C, McIntosh S, Hassan AE, Matsumoto K (2017b) An Empirical Comparison of Model Validation Techniques for Defect Prediction Models. Transactions on Software Engineering (TSE) 43(1):1–18

Thongtanunam P, McIntosh S, Hassan AE, Iida H (2015a) Investigating Code Review Practices in Defective Files: An Empirical Study of the Qt System. In: Proceedings of the 12th Working Conference on Mining Software Repositories (MSR), pp 168–179

Thongtanunam P, Tantithamthavorn C, Kula RG, Yoshida N, Iida H, Matsumoto K (2015b) Who Should Review My Code? A File Location-Based Code-Reviewer Recommendation Approach for Modern Code Review. In: Proceedings of the the 22nd International Conference on Software Analysis, Evolution, and Reengineering (SANER), pp 141–150

Thongtanunam P, McIntosh S, Hassan AE, Iida H (2016a) Review Participation in Modern Code Review: An Empirical Study of the Android, Qt, and OpenStack Projects. Empirical Software Engineering (EMSE) 22(2):768–817

Thongtanunam P, McIntosh S, Hassan AE, Iida H (2016b) Revisiting Code Ownership and Its Relationship with Software Quality in the Scope of Modern Code Review. In: Proceedings of the 38th International Conference on Software Engineering (ICSE), pp 1039–1050

Vasilescu B, Serebrenik A, Devanbu P, Filkov V (2014) How Social Q&A Sites are Changing Knowledge Sharing in Open Source Software Communities. In: Proceedings of the 17th ACM

Conference on Computer Supported Cooperative Work & Social Computing (CSCW), pp 342–354

van Wesel P, Lin B, Robles G, Serebrenik A (2017) Reviewing Career Paths of the OpenStack Developers. In: Proceedings of the 33rd International Conference on Software Maintenance and Evolution (ICSME), pp 544–548

Whitehead J (2007) Collaboration in Software Engineering: A Roadmap. In: Proceedings of the 2007 Future of Software Engineering (FOSE), pp 214–225

Xia X, Lo D, Wang X, Yang X (2015) Who Should Review This Change?: Putting Text and File Location Analyses Together for More Accurate Recommendations. In: Proceedings of the 31st International Conference on Software Maintenance and Evolution (ICSME), pp 261–270

Yang X, Kula RG, Yoshida N, Iida H (2016a) Mining the Modern Code Review Repositories: A Dataset of People, Process and Product. In: Proceedings of the 13th International Conference on Mining Software Repositories (MSR), pp 460–463

Yang X, Yoshida N, Kula RG, Iida H (2016b) Peer Review Social Network (PeRSoN) in Open Source Projects. Transactions on Information and Systems E99.D(3):661–670

Yu Y, Wang H, Yin G, Ling CX (2014) Reviewer Recommender of Pull-Requests in GitHub. In: Proceedings of the 30th International Conference on Software Maintenance and Evolution (ICSME), pp 610–613

Zanjani MB, Kagdi H, Bird C (2016) Automatically Recommending Peer Reviewers in Modern Code Review. Transactions on Software Engineering (TSE) 42(6):530–543

Zimmermann T, Zeller A, Weissgerber P, Diehl S (2005) Mining Version Histories to Guide Software Changes. Transactions on Software Engineering (TSE) 31(6):429–445